\documentclass[aps,twocolumn]{revtex4}

\usepackage{graphicx}
\usepackage{bm}
\usepackage{braket}
\usepackage{hyperref}
\usepackage[usenames]{xcolor}
\usepackage{amsmath}
\usepackage{graphicx}
\usepackage{latexsym}
\usepackage{amsmath,amssymb}
\usepackage{xcolor}
\usepackage{stackrel}
\usepackage{accents}
\usepackage{empheq}
\usepackage[title]{appendix}

\DeclareMathOperator{\sgn}{sgn}

\newcommand{\im}{\operatorname{Im}}
\newcommand{\re}{\operatorname{Re}}

\newcommand{\vex}[1]{\bm{\mathrm{#1}}}
\newcommand{\msf}[1]{\mathsf{#1}}

\newcommand{\dd}{\mathcal{D}}
\newcommand{\intl}[1]{\int\limits_{#1}}
\newcommand{\suml}[1]{\sum\limits_{#1}}

\newcommand{\kb}{\vex{k}}
\newcommand{\pb}{\vex{k}}
\newcommand{\qb}{\vex{q}}
\newcommand{\Qb}{\vex{Q}}
\newcommand{\rb}{\vex{r}}
\newcommand{\xb}{\vex{x}}
\newcommand{\ww}{\omega}
\newcommand{\WW}{\Omega}
\newcommand{\e}{\varepsilon}

\newcommand{\htau}{\tau}

\newcommand{\bpsi}{\bar{\psi}}

\newcommand{\pf}{k_{\msf{F}}}
\newcommand{\vf}{v_{\msf{F}}}

\newcommand{\Tf}{E_{\msf{F}}}
\newcommand{\ktf}{\kappa}

\newcommand{\GG}{\mathcal{G}}

\newcommand{\G}{G}
\newcommand{\T}{\mathsf{T}}
\newcommand{\D}{\mathcal{D}}

\newcommand{\tim}{\mathrm{T}}
\newcommand{\atim}{\bar{\mathrm{T}}}

\newcommand{\mf}{M_F}
\newcommand{\uk}{U_{\msf{K}}}

\newcommand{\phicl}{\phi_{\mathsf{cl}}}
\newcommand{\phiq}{\phi_{\mathsf{q}}}
\newcommand{\Jcl}{J_{\mathsf{cl}}}
\newcommand{\Jq}{J_{\mathsf{q}}}

\newcommand{\LL}{\msf{L}}
\newcommand{\RR}{\msf{R}}
\newcommand{\IL}{1_{\msf{L}}}
\newcommand{\IS}{1_{\sigma}}
\newcommand{\Txy}{T_{xy}}
\newcommand{\tF}{\tilde{F}}
\newcommand{\rs}{r_{\msf{s}}}

\begin{document}

\title{Drag viscosity of metals and its connection to Coulomb drag }

\author{Yunxiang Liao}
\author{Victor Galitski}%
\affiliation{Joint Quantum Institute, University of Maryland, College Park, Maryland, 20742, USA}
\affiliation{Condensed Matter Theory Center, University of Maryland, College Park, MD 20742, USA}

\date{\today}

\begin{abstract}
Recent years have seen a surge of interest in studies of hydrodynamic transport in electronic systems. 
We investigate the electron viscosity of metals and find a new component that is closely related to Coulomb drag.
Using the linear response theory, viscosity, a transport coefficient for momentum, can be extracted from the retarded correlation function of the momentum flux, i.e., the stress tensor. There exists a previously overlooked contribution to the shear viscosity from the interacting part of the stress tensor which accounts for the momentum flow induced by interactions. This contribution, which we dub drag viscosity, is caused by the frictional drag force due to long-range interactions. It is therefore linked to Coulomb drag which also originates from the interaction induced drag force. Starting from the Kubo formula and using the Keldysh technique, we compute the drag viscosity of 2D and 3D metals along with the drag resistivity of double-layer 2D electronic systems. Both the drag resistivity and drag viscosity exhibit a crossover from quadratic-in-$T$ behavior at low temperatures to a linear one at higher temperatures. 
Although the drag viscosity appears relatively small compared with the normal Drude component for the clean metals, it may dominate hydrodynamic transport in some systems, which are discussed in the conclusion.

\end{abstract}           
 
\maketitle



\section{Introduction}

Hydrodynamics is an effective description of dynamics of system at length and time scales large compared with that of local equilibration processes.
It is governed by the conservation laws, and the hydrodynamics equations can be derived from the continuity equations along with the constitutive relations which connect the conserved currents with the hydrodynamics variables.
It is  useful to a wide variety of systems,
but is usually not relevant to electron liquid in solids due to momentum relaxing processes such as electron-impurity, Umklapp, or electron-phonon scattering.
However, if the dominant scattering process is due to electron-electron interactions, the electron liquid can be described hydrodynamically as it reaches local equilibrium at the scale of electron-electron scattering length which is much smaller compared with that of momentum relaxing processes.
In recent years, hydrodynamic behavior of electrons has attracted a lot of interest~\cite{ElHydro,GHydro-1,GHydro-2,GHydro-3,Muller,GrapheneV-1,GrapheneV-2,GrapheneV-3,GrapheneV-4,Hall,Hall-2,EH-1, EH-2, EH-3, EH-4,EH-5, EH-6,  Apostolov,Apostolov2,VS}, and has been observed experimentally in graphene~\cite{GH-Kumar,GH-Bandurin,GH-Crossno}, Weyl semimetals~\cite{Weyl,WP2}, ultrapure PdCoO$_2$~\cite{PdCoO2}, PtSn$_4$~\cite{PtSn4} and GaAs~\cite{GaAs}

In the hydrodynamics equations, the underlying microscopic physics is encoded by transport coefficients which measure a fluid's resistance to velocity or thermal gradient. Viscosity is a transport coefficient for momentum and enters the Navier-Stokes equation. For a fluid with inhomogeneous flow velocity, the transverse velocity gradient induces a friction force which acts between the adjacent layers of fluid and resists their relative motion. The shear viscosity is defined as the ratio of induced force per unit area over the velocity gradient, see Ref.~\cite{VisRev} for a review. 
There exists another momentum transport coefficient named bulk viscosity which also enters the Navier-Stokes equation but will not be considered in the present paper.
The ratio of shear viscosity to entropy density has been considered as a measure of fluid's interaction strength, with a lower bound conjectured by Kovtun et al.~\cite{KSS}. 
Using linear response theory,
viscosity can be extracted from the Kubo formula which relates viscosity with the correlation function of momentum flux, or stress tensor (see Sec.~\ref{sec:Kubo} for details).

The viscosity of a 3D Fermi liquid was first studied by Pomeranchuk~\cite{pomeranchuk} who  found a $1/T^2$ temperature dependence through the dimensional analysis.
This result was later confirmed by Abrikosov and Khalatnikov~\cite{Abrikosov,abrikosov1957} using the kinetic equation approach, and is in good agreement with the $^3$He experiment (see review~\cite{He3}).
It can be recovered from a ``Drude''-type diagram which corresponds to the correlation function of the noninteracting part of stress tensor~\cite{VS}.

As emphasized by Nishida and Son~\cite{Nishida} as well as Link et al.~\cite{Schmalian}, there exists an additional contribution to shear viscosity from the interacting part of the stress tensor.
As shown in the present paper, this contribution is related to the interaction induced drag force, and will be called drag viscosity in the following.
Interactions between two distant layers of fluid with different transverse velocities give rise to the drag force that acts between the two layers and tends to reduce their relative motion, and the corresponding contribution to shear viscosity is the drag viscosity.

The frictional force giving rise to drag viscosity is also responsible for Coulomb drag,
which was first proposed by Pogrebinskii~\cite{pogrebinskii} and later by Price~\cite{Price1983,Price1988} in an attempt to separate electron-electron scattering from other scattering processes and to study the direct effect of interactions on the transport properties.
In a standard setup, one considers two closely spaced electronic layers, to one of which a current $I_{\LL}$ is applied. Interlayer interactions induce a drag force which acts on the electrons of the second layer and drags them along.  If no current is allowed to flow in that layer, this leads to an induced voltage $V_{\RR}$ such that the resulting electric force cancels with the drag force. The ratio of the induced voltage $V_{\RR}$ to the applied current $I_{\LL}$ is known as the drag resistance
$
R_{\msf{D}}=V_{\RR}/I_{\LL}.
$

Since the pioneering experiment by Gramila et al.~\cite{Gramila} on Coulomb drag of a double-layer 2D electron gas (2DEG) system, there have been extensive studies~\cite{MacDonald,Flensberg,Kamenev,Solomon,Jauho,Langevin,Flensberg,Flensberg2,plasmon,Langevin,Apostolov,Apostolov,Width,XC} investigating the underlying mechanism~(see reviews~\cite{DragRev,Rojo} and references therein).
It has been found that, in the ballistic limit and in the temperature region where the plasmon contribution is unimportant, the temperature dependence of drag resistivity crosses over from quadratic in $T$~\cite{MacDonald,Flensberg,Kamenev} to linear in $T$~\cite{Solomon,Jauho,Langevin} as $T$ increases.
The plasmon contribution becomes important at higher temperature leading to a deviation from the linear behavior~\cite{Flensberg,Flensberg2,plasmon,Langevin}.
For strongly correlated systems with large dimensionless interaction parameter $\rs \gg 1$, a hydrodynamics approach has been employed by Apostolov et al.~\cite{Apostolov,Apostolov2} to study the drag resistivity which is then expressed in terms transport coefficients such as viscosity.

In this paper, we study drag viscosity along with Coulomb drag, both of which result from the interaction induced drag force, with our emphasis placed on the close resemblance between them. 
Using the linear response theory and the Keldysh technique, we compute the drag viscosity, which remains overlooked in the literature.  In addition, we derive in an analogous manner the Coulomb drag in the clean limit, and recover the results obtained using different theoretical approaches from previous studies~\cite{MacDonald,Flensberg, Kamenev,Solomon,Jauho,Langevin}. We compare the two quantities and find that they exhibit the same temperature dependence with different values of the crossover temperature.

The rest of the paper is organized as follows. 
In Sec.~\ref{sec:Kubo}, we present the Kubo formula connecting the correlation function of the interacting stress tensor with drag viscosity, and compare it with the one connecting the correlation function of the drag force with drag resistivity.
The derivation of the interacting part of the stress tensor, which describes the interaction induced momentum flow, is also presented in this section.
Starting from the Kubo formula and using the Keldysh technique, in Sec.~\ref{sec:Derivation}, we derive the general expressions for the drag viscosity of clean electronic systems in 2D and 3D along with the drag resistivity of double-layer 2DEGs. The latter is consistent with the result previously obtained in Ref.~\cite{Langevin} using the Boltzmann-Langevin kinetic theory. 
In Sec.~\ref{sec:Results}, we present and analyze the results in different temperature regions.
Finally, in Sec.~\ref{sec:conclusion}, we conclude with a brief summary together with a discussion of open questions and directions for future studies.	
In Appendix~\ref{app:diagram}, we provide an alternative diagrammatic approach for the derivation of the correlation function of the interacting stress tensor and that of the drag force.
In Appendix~\ref{app:Pi}, the explicit expressions for the polarization operator from Refs.~\cite{stern,lindhard} are presented.


\section{Kubo Formulas for Drag Viscosity and Drag Resistivity}~\label{sec:Kubo}

\subsection{Drag Viscosity and Interacting Stress Tensor}

Hydrodynamics is governed by the conservation of particle number, energy and momentum. The dissipation which affects the flux of these conserved quantities is described in terms of the transport coefficients. One of them is the viscosity which measures the rate of momentum transport. The corresponding conservation equation, i.e., the continuity equation for momentum is given by
\begin{align}\label{eq:Pcon}
	\partial_t g_{\alpha}(\rb,t)
	+
	\partial_{\beta} T_{\alpha \beta}(\rb,t)
	=0.
\end{align}
Here, $g_{\alpha}(\rb,t)$ represents the momentum density at position $\rb$ and time $t$ in the $\alpha$-direction, and $T_{\alpha \beta}(\rb,t)$ denotes the stress tensor (or the momentum flux).
We have employed the convention of summation over repeated indices.

The equation above suggests that the stress tensor is essential to the calculation of viscosity. Using the linear response theory, it has been found that the shear viscosity $\eta$ can be obtained from the following Kubo's formula~\cite{Kadanoff,Read,forster,zubarev},
\begin{align}\label{eq:Kubo-V}
\begin{aligned}
\eta
=\,&
\lim\limits_{\omega \rightarrow 0}
\lim\limits_{\qb \rightarrow \vex{0}}
\frac{-1}{\omega}
\operatorname{Im}
\GG^{(R)}_{T}(\qb,\omega),	 
\end{aligned}
\end{align}
where $\GG^{(R)}_{T}(\qb,\omega)$ indicates the retarded correlation function of the stress tensor $\Txy$,
\begin{align}\label{eq:GT}
\begin{aligned}
	\GG^{(R)}_{T}(\qb,\omega)
	=\,&
	-i
	\int_{-\infty}^{+\infty}dt
	\int d^{D}\rb
	\,
	e^{i\omega t - i \qb \cdot \rb}
	\Theta(t)
	\\
	\times&
	\braket{
		\left[ 
		\Txy (\rb,t),
		\Txy (\vex{0},0)
		\right] 
	}.
\end{aligned}
\end{align}

The stress tensor is composed of a non-interacting part $T^{(0)}_{\alpha \beta}$ accounting for the single-particle contribution to the momentum flow, and an interacting part $T^{(\msf{int})}_{\alpha \beta}$ associated with the momentum transport induced by interactions.
The latter contribution to shear viscosity is absent in the case of contact interactions and is usually overlooked in previous studies. 
In this section, we present a brief derivation of the interacting part of the stress tensor. Only its zero momentum Fourier component (spatial average) $T_{\alpha \beta}(\Qb=0) = \int_{\rb} \T_{\alpha \beta}(\rb)$, which enters the Kubo's formula for shear viscosity $\eta$ (Eqs.~\ref{eq:Kubo-V} and~\ref{eq:GT}), is derived here. See Ref.~\cite{Schwinger} for the complete expression of the stress tensor and the corresponding derivation, and Ref.~\cite{geoT} for an alternative approach.

Consider a fermionic system described by the following Hamiltonian,
\begin{align}\label{eq:H}
\begin{aligned}
	&H=\,
	\intl{\rb}
	\frac{1}{2m}
	\bm{\nabla} \psi^{\dagger}_{\sigma}  (\rb,t)
	\cdot
	\bm{\nabla} \psi_{\sigma}  (\rb,t)
	\\
	+&
	\frac{1}{2} 
	\intl{\rb,\rb'}
	\psi^{\dagger}_{\sigma} (\rb,t)
	\psi^{\dagger}_{\sigma'}  (\rb',t)
	V(\rb-\rb')
	\psi_{\sigma'}  (\rb',t)
	\psi_{\sigma}  (\rb,t),
\end{aligned}	
\end{align}
where $\psi_{\sigma}$ indicates the fermion with spin $\sigma$, and $V( \rb-\rb' )= V(\lvert \rb-\rb' \rvert)$ is the interaction potential. Throughout this paper, we work with the units $\hbar=k_{\msf{B}}=1$.

The momentum density $\vex{g}(\rb,t)$ is given by
\begin{align}\label{eq:G}
\begin{aligned}
	g_{\alpha}(\rb,t)
	=\,
	\frac{1}{2i}
	\left[ 
	\psi^{\dagger}_{\sigma} (\rb,t)
	\partial_{\alpha}
	\psi_{\sigma} (\rb,t)
	-
	\partial_{\alpha}
	\psi^{\dagger}_{\sigma} (\rb,t)
	\psi_{\sigma} (\rb,t)
	\right].
\end{aligned}	
\end{align}
Using the equation of motion 
\begin{align}
\begin{aligned}
	&i \frac{\partial}{\partial t} \psi_{\sigma} (\rb, t)
	=\,
	-\frac{1}{2m}
	\nabla^2 \psi_{\sigma}  (\rb,t)
	\\
	&+
	\intl{\rb'}
	\psi^{\dagger}_{\sigma'}  (\rb',t)
	V(\rb-\rb')
	\psi_{\sigma'}  (\rb',t)
	\psi_{\sigma}  (\rb,t),
\end{aligned}	
\end{align}
the time derivative of the momentum density can be expressed as
\begin{align}\label{eq:dGdt}
\begin{aligned}
	&\frac{\partial g_{\alpha}(\rb,t)}{\partial t}
	=\,
	-\partial_{\beta}T^{(0)}_{\alpha \beta}(\rb,t)
	\\
	&
	-
	\intl{\rb'}
	\psi^{\dagger}_{\sigma} (\rb,t)
	\psi^{\dagger}_{\sigma'}  (\rb',t)
	\frac{\partial V(\rb-\rb')}{\partial {r_\alpha}}
	\psi_{\sigma'}  (\rb',t)
	\psi_{\sigma}  (\rb,t),
\end{aligned}	
\end{align}
where
\begin{align}\label{eq:T0}
\begin{aligned}
&T^{(0)}_{\alpha \beta}(\rb,t)
=
\frac{1}{2m}
\left\lbrace 
\partial_\alpha \psi^{\dagger}_{\sigma} (\rb,t)
\partial_\beta \psi_{\sigma} (\rb,t)
\right. 
\\
&\left. 
+
\partial_\beta \psi^{\dagger}_{\sigma} (\rb,t)
\partial_\alpha \psi_{\sigma} (\rb,t)
-\frac{1}{2}
\delta_{\alpha\beta} \nabla^2 
\left[  \psi^{\dagger}_{\sigma}(\rb,t)  \psi_{\sigma}(\rb,t) \right] 
\right\rbrace .
\end{aligned}
\end{align}
Here the first term on the right-hand side of Eq.~\ref{eq:dGdt} arises from the non-interacting part of the Hamiltonian, whereas the second term is from the interacting part.

From the continuity equation for momentum (Eq.~\ref{eq:Pcon}), one can deduce that the non-interacting part of the stress tensor is given by $T^{(0)}_{\alpha \beta}(\rb,t)$ defined in Eq.~\ref{eq:T0}.
The interacting part $T^{(\msf{int})}_{\alpha \beta}(\rb,t)$ is subject to the following condition
\begin{align}
\begin{aligned}
&\partial_\beta T^{(\msf{int})}_{\alpha \beta}(\rb,t)
\\
=&
\intl{\rb'}
\psi^{\dagger}_{\sigma} (\rb,t)
\psi^{\dagger}_{\sigma'}  (\rb',t)
\frac{\partial V(\rb-\rb')}{\partial r_{\alpha}}
\psi_{\sigma'}  (\rb',t)
\psi_{\sigma}  (\rb,t),
\end{aligned}
\end{align}
whose Fourier transform is 
\begin{align}\label{eq:T1EOM}
\begin{aligned}
Q_{\beta} T^{(\msf{int})}_{\alpha \beta}(\Qb,t)
=\,&
\intl{\pb,\pb',\qb}
\psi^{\dagger}_{\sigma} (\pb+\qb-\Qb,t)
\psi^{\dagger}_{\sigma'}  (\pb'-\qb,t)
\\
\times &
q_{\alpha}
V(q)
\psi_{\sigma'}  (\pb',t)
\psi_{\sigma}  (\pb,t).
\end{aligned}
\end{align}
Here we have used the shorthand notation $\int_{\kb}\equiv \int \frac{d^D k}{(2\pi)^D}$ with $D$ being the spatial dimensions.

In the calculation of shear viscosity, only the zero momentum Fourier component of the stress tensor $T_{xy}(\Qb=0)$ is needed (see Kubo's formula Eqs.~\ref{eq:Kubo-V} and~\ref{eq:GT}).
We thus expand the right-hand side of Eq.~\ref{eq:T1EOM} around $\Qb=0$. At the zeroth order in $\Qb$, the integrand is an odd function of $\qb$ and vanishes after integration.
The first order term is given by
\begin{align}\label{eq:Ok1}
\begin{aligned}
	&
	-Q_{\beta}
	\intl{\pb,\pb',\qb}
	 \partial_{q_{\beta}} \psi^{\dagger}_{\sigma} (\pb+\qb) 
	\psi^{\dagger}_{\sigma'}  (\pb'-\qb)
	q_{\alpha} V(q)
	\psi_{\sigma'}  (\pb')
	\psi_{\sigma}  (\pb)	
	\\
	&=\,
	Q_{\beta}
	\intl{\pb,\pb',\qb}
	\psi^{\dagger}_{\sigma} (\pb+\qb) 
	 \partial_{q_{\beta}}\psi^{\dagger}_{\sigma'}  (\pb'-\qb)
	q_{\alpha} V(q)
	\psi_{\sigma'}  (\pb')
	\psi_{\sigma}  (\pb)	
	\\
	&
	+Q_{\beta}
	\intl{\pb,\pb',\qb}
	\psi^{\dagger}_{\sigma} (\pb+\qb) 
	\psi^{\dagger}_{\sigma'}  (\pb'-\qb)
	\partial_{q_{\beta}} \left[ q_{\alpha} V(q) \right] 
	\psi_{\sigma'}  (\pb')
	\psi_{\sigma}  (\pb),
\end{aligned}
\end{align}	
where we have suppressed the time argument of the fermionic field $\psi$ for brevity, and applied an integration by parts.
Through a change of variables, it is easy to prove that the first term on the right-hand side of the equation above can be rewritten as 
\begin{align}\label{eq:Ok2}
\begin{aligned}
Q_{\beta}
\intl{\pb,\pb',\qb}
 \partial_{q_{\beta}}\psi^{\dagger}_{\sigma} (\pb+\qb)
\psi^{\dagger}_{\sigma'}  (\pb'-\qb)
q_{\alpha} V(q)
\psi_{\sigma'}  (\pb')
\psi_{\sigma}  (\pb),	
\end{aligned}
\end{align}
which equals the negative of the left-hand side of Eq.~\ref{eq:Ok1}.	

Using Eqs.~\ref{eq:Ok1} and~\ref{eq:Ok2}, one obtains the first order term in $\Qb$ on the right-hand side of Eq.~\ref{eq:T1EOM} 
\begin{align}
\begin{aligned}
\frac{1}{2}
Q_{\beta}
\intl{\pb,\pb',\qb}
\psi^{\dagger}_{\sigma} (\pb+\qb) 
\psi^{\dagger}_{\sigma'}  (\pb'-\qb)
\partial_{q_{\beta}} \left[q_{\alpha} V(q) \right]
\psi_{\sigma'}  (\pb')
\psi_{\sigma}  (\pb).
\end{aligned}
\end{align}		
Comparing it with the left-hand side of the same equation, we arrive at the zero momentum Fourier component of the interacting stress tensor
\begin{align}\label{eq:Tint}
\begin{aligned}
	T_{\alpha \beta}^{\msf{(int)}}(\vex{0},t)
	=\,&
	\frac{1}{2}
	\intl{\pb,\pb',\qb}
	\left[  \delta_{\alpha\beta}V(q)+\frac{q_\alpha q_\beta }{q} V'(q) \right] 
	\\
	\times &
	\psi^{\dagger}_{\sigma} (\pb+\qb,t) 
	\psi^{\dagger}_{\sigma'}  (\pb'-\qb,t)
	\psi_{\sigma'}  (\pb',t)
	\psi_{\sigma}  (\pb,t).
\end{aligned}
\end{align}	
From this result, one can see that the off-diagonal component ($\alpha \neq \beta$) of the interacting part of the stress tensor $T^{(\msf{int})}_{\alpha \beta}$ is vanishing for contact interactions since $V'(q)=0$.
We note that the expression for interacting stress tensor only depends on the interacting part of the Hamiltonian (Eq.~\ref{eq:H}) and is therefore valid for dispersion other than parabolic.

In this paper, we investigate the contribution to the shear viscosity from the interacting stress tensor $T^{(\msf{int})}_{xy}$. In particular, we focus on the part associated with the correlation function of $T^{(\msf{int})}_{xy}$, which will be called the drag viscosity $\eta_{\msf{D}}$ in the following,
\begin{align}~\label{eq:etaD}
	\eta_{\msf{D}}
	=\lim\limits_{\ww\rightarrow 0}
	\frac{1}{\ww}\im 
	\intl{\rb,t}
	e^{i\ww t}i \Theta (t)
	\braket{\left[ T^{(\msf{int})}_{xy}(\rb,t),T^{(\msf{int})}_{xy}(\vex{0},0)\right]  }.
\end{align}
There is an additional contribution from the correlation function of the non-interacting and interacting parts of the stress tensor $\braket{\left[ T^{(0)}(\rb,t),T^{(\msf{int})}(\vex{0},0)\right]  }$, which has been thoroughly studied in Ref.~\cite{Schmalian} for intrinsic graphene in the collisionless regime at zero temperature.
Here we ignore this contribution as our emphasis is placed on the similarity between the drag viscosity $\eta_{\msf{D}}$ and Coulomb drag.
Furthermore, for some systems with unscreened long-range interactions, due to the appearance of the derivative of interaction potential $V’(q)$ in the stress tensor formula (Eq. 14), the contribution from the drag viscosity may be dominant, and is the focus of the present paper.

\subsection{Drag Resistivity and Drag Force}

The drag viscosity discussed in the previous subsection measures the transverse momentum transport induced by long-range interactions. 
It is closely related to Coulomb drag which arises from momentum transfer between two electronic layers in the presence of interlayer Coulomb scatterings.

In Ref.~\cite{MacDonald}, Zheng and MacDonald employ the memory function formalism to investigate Coulomb drag, and convert the Kubo formula relating the current-current correlation function and drag conductivity into the one which connects the force-force correlation function and drag resistivity:
\begin{align}\label{eq:Kubo-R}
\begin{aligned}
	\rho_{\msf{D}}
	=\,&
	-\rho_{xx}^{\LL\RR}
	=
	\frac{1}{n_L n_R e^2}
	\lim\limits_{\omega \rightarrow 0}
	\frac{1}{\omega}
	\operatorname{Im}
	\GG^{(R)}_F(\ww),
	\\
	\GG^{(R)}_F(\ww)
	=\,&
	-i
	\int_{-\infty}^{+\infty}dt
	\,
	e^{i\omega t }
	\Theta(t)
	\braket{
		\left[ 
		F^{\LL}_x(t),
		F^{\RR}_x(0)
		\right] 
	}.
\end{aligned}
\end{align}
Here the drag force is defined as
\begin{align}\label{eq:F}
\begin{aligned}
	&\vex{F}^{\LL}(t)
	=
	-\vex{F}^{\RR}(t)
	=
	\intl{\kb,\kb',\qb} i \vex{q} V_{\LL \RR}(q)
	\\
	\times &
	\psi^{\dagger}_{\sigma,\LL} (\kb+\qb,t)
	\psi^{\dagger}_{\sigma',\RR}  (\kb'-\qb,t)
	\psi_{\sigma',\RR}  (\kb',t)
	\psi_{\sigma,\LL}  (\kb,t).
\end{aligned}
\end{align}
$\LL$ and $\RR$ denote the layer indices.
$V_{\LL \RR}(q)$ is the interlayer interaction potential. 
$n_{\LL}$ ($n_{\RR}$) indicates the density of electrons in layer $\LL$ ($\RR$), while $\vex{F}^{\LL}$ ($\vex{F}^{\RR}$) represents the frictional drag force acting on the electrons in layer $\LL$ ($\RR$).

The drag force $F$ (Eq.~\ref{eq:F}) which enters the Kubo's formula for drag resistivity acquires a form analogous to that of interacting stress tensor $\Txy^{(\msf{int})}$ (Eq.~\ref{eq:Tint}) whose correlation function gives the drag viscosity (Eq.~\ref{eq:Kubo-V}). In the following, we calculate the drag viscosity and drag resistivity in parallel and show the close resemblance between the two.  

\section{Derivation of Drag Viscosity and Drag Resistivity}~\label{sec:Derivation}

\subsection{Keldysh Formalism}

In this section, we derive general expressions for drag viscosity and the drag resistivity expressed in terms of polarization operator $\Pi$ and bare interaction potential $V$.
The starting point is the partition function in the Keldysh formalism
\begin{align}\label{eq:ZK}
\begin{aligned}
	Z
	=\,&
	\int 
	\D \left( \bar{\psi}, \psi \right) 
	\,
	\exp
	\left( 
	iS_0+iS_{\msf{int}}
	\right) ,
	\\
	iS_0[\psi]
	=\,&
	i
	\intl{\vex{x},\vex{x'},t,t'}
	\bar{\psi}(\vex{x},t)
	\;
	\G_0^{-1}(\vex{x},t;\vex{x'},t')	
	\;
	\psi(\vex{x'},t'),
	\\
	iS_{\msf{int}}[\psi]
	=\,&
	-
	{\frac{i}{2}}
	\,
	\suml{a=1,2}
	\suml{i,j=\RR,\LL}
	\zeta_a
	\intl{t,\vex{x}}
	V_{ij}(\xb-\xb')
	\\
	\times &
	\bar{\psi}^{i}_{a,\sigma}(\xb,t) \bar{\psi}^{j}_{a,\sigma'}(\xb',t)
	\psi^{j}_{a,\sigma'}(\xb',t)\psi^{i}_{a,\sigma}(\xb,t).
\end{aligned}	
\end{align}
Here the fermionic field $\psi$ resides on a closed Keldysh contour that goes from $t=-\infty$ to $t=+\infty$ and back to $t=-\infty$, and its index $a=1$ ($a=2$) corresponds to the forward (backward) part of the contour. 
$\psi$ carries an additional spin index $\sigma \in\left\lbrace \uparrow, \downarrow \right\rbrace $, as well as a layer index $i \in \left\lbrace \LL,\RR \right\rbrace $ for the double-layer system considered in the calculation of drag resistivity.
As for the drag viscosity problem, the layer index $i$ and the summation over it should be ignored. 
$\zeta_a$ is a sign factor that takes the value of $+1$ ($-1$) for $a=1$ ($a=2$).

$\G_0$ stands for the non-interacting single-particle Green's function defined on the Keldysh contour: 
\begin{align}
	G_0(\xb,t;\xb',t')
	=
	-i \left\langle \tim_{\msf{c}} \, \psi(\xb,t) \, \bar{\psi}(\xb',t')\right\rangle_0.
\end{align}
Here $\tim_{\msf{c}}$ denotes the contour ordering operator, and the angular bracket with subscript 0 indicates functional averaging that does not include the interactions.
In the Keldysh space, $\G_0$ assumes the following structure:
\begin{align}
	\G_0
	\equiv
	\begin{bmatrix}
	\G_0^{(\tim)}
	&
	\G_0^{(<)}
	\\
	\G_0^{(>)}
	&
	\G_0^{(\atim)}
	\end{bmatrix},
\end{align}
where $	\G_0^{(\tim)}$, $\G_0^{(\atim)}$, $\G_0^{(<)}$ and $\G_0^{(>)}$ represent, respectively, the time-ordered, anti-time-ordered, lesser and greater Green's functions.

The interaction potential $V(q)$ is given by $V(q)=4 \pi e^2/q^2$ $\left[ V(q)=2 \pi e^2/q\right] $ for the calculation of viscosity of a 3D (2D) electronic system.
On the other hand, for drag resistivity of a double-layer 2DEG, $V(q)$ in the layer space takes the form of
\begin{align}\label{eq:V}
	V(q)
	=
	\begin{bmatrix}
	V_{\msf{intra}}(q) & V_{\msf{inter}}(q)
	\\
	V_{\msf{inter}}(q) & V_{\msf{intra}}(q)
	\end{bmatrix},
\end{align}
where $V_{\msf{intra}}(q)=2 \pi e^2 /q$ [$V_{\msf{inter}}(q)=2 \pi e^2 e^{-qd}/q$] denotes the intralayer (interlayer) interaction, with $d$ being the interlayer distance.

It is convenient to introduce a Hubbard-Stratonovich (H.S.) decoupling of the interaction:
\begin{align}
\begin{aligned}
	&
	e^{iS_{\msf{int}}}
	=
	\int 
	\D \phi
	\exp
	\left[
	\frac{i}{2} \sum_{a }
	\zeta_a
	\intl{\qb,\ww}
	\phi^{i}_a(\qb,\ww) V^{-1}_{ij}(q)  \phi^{j}_a(-\qb,-\ww)
	\right. 
	\\
	&
	\left. 
	-i
	\sum_{a,i}
	\zeta_a
	\intl{\e,\kb}\intl{\ww,\qb}
	\phi^{i}_{a}(\qb,\ww) 
	\bar{\psi}^{i}_{a,\sigma}(\kb+\qb,\e+\ww)
	\psi^{i}_{a,\sigma}(\kb,\e) 
	\right].
\end{aligned}
\end{align}
Here $\phi^{i}_{a}$ is a bosonic field that carries a Keldysh index $a$ and a layer index $i$,
and $\int_{\ww}$ is a shorthand for $\int_{-\infty}^{\infty} d\ww/2\pi$.

To further simplify the calculation, we apply a Keldysh rotation for the H.S. decoupling field $\phi$:
\begin{align}\label{eq:KRot-phi}
\phicl
\equiv
\left( \phi_1+\phi_2\right) /\sqrt{2},
\qquad
\phiq
\equiv
\left( \phi_1-\phi_2\right) /\sqrt{2},
\end{align}
as well as a nonunitary transformation to the fermionic field $\psi$ in the Keldysh space:
\begin{align}\label{eq:CoV}
\begin{aligned}
&\psi(\kb,\e) 
\rightarrow \, 
\htau^3 \uk \mf(\e) \, \psi(\kb,\e),
\\
&\bar{\psi}(\kb,\e) 
\rightarrow \,
\bar{\psi}(\kb,\e) \, \mf(\e) \uk^\dagger.
\end{aligned}
\end{align} 
Here $\htau$, $\uk$ and $\mf(\e)$ are matrices that only act on the Keldysh space. $\htau$ stands for the Pauli matrix in the Keldysh space, while $\uk$ and $\mf(\e)$ are defined as
\begin{align}
\begin{aligned}
	\uk \equiv
	{\textstyle{\frac{1}{\sqrt{2}}}}
	(1 + i \htau^2),
	\,\,
	\mf(\e)
	\equiv
	\begin{bmatrix}
	1  	& \tanh \left(\frac{ \e}{2 T}\right)
		\\
	0 	& -1  
	\end{bmatrix}.
\end{aligned}
\end{align}

The transformation defined by Eq~\ref{eq:CoV} is a successive application of a Keldysh rotation and a thermal rotation which depends on the generalized fermionic distribution function $\tanh \left(\frac{ \e}{2 T}\right)$. 
After this transformation, the non-interacting fermionic Green's function becomes distribution function independent and diagonal in the Keldysh space:
\begin{align}
	G'_0=\,
	\begin{bmatrix}
	G_0^{(R)} & 0
	\\
	0 & G_0^{(A)}
	\end{bmatrix}
	\otimes
	\IS
	\otimes
	\IL.
\end{align}
Here $\IS$ and $\IL$ represent, respectively, identical matrices in the spin and layer spaces. $G_0^{(R/A)}$ is the non-interacting retarded (advanced) Green's function for the fermionic field $\psi$ and takes the from
\begin{align}\label{eq:G0}
	G_0^{(R/A)}(\kb,\e)
	=
	\left[  \e-\xi_{\kb}\pm i\eta\right]^{-1},
\end{align}
where $\xi_{\kb} \equiv k^2/2m-\mu$ with $\mu$ being the chemical potential, and $\eta$ is a positive infinitesimal. 
Under the transformations Eqs~\ref{eq:KRot-phi} and~\ref{eq:CoV}, the partition function changes to
\begin{widetext}
	\begin{subequations}\label{eq:Z-rot}
	\begin{align}
	\label{eq:Z}
	&
	Z
	=
	\int 
	\D \left( \bar{\psi}, \psi \right) 
	\D \phi
	\exp\left( iS_{\phi}+iS_{\psi}+iS_c\right), 
	\\
	\label{eq:Sphi}
	&S_{\phi}[\phi]
	=
	\intl{\qb,\ww}
	\phicl(\qb,\ww) V^{-1}(q) \phiq(-\qb,-\ww) ,
	\\
	\label{eq:Spsi}
	& S_{\psi}[\bpsi,\psi]
	=\,	
	\intl{\kb,\e}
	\bar{\psi}(\kb,\e)
	G'_{0}\,^{-1}(\kb,\e)
	\psi(\kb,\e),
	\\
	\label{eq:Sc}
	&\begin{aligned}
	&S_c[\phi,\bpsi,\psi]
	=
	-
	\frac{1}{\sqrt{2}} 
	\intl{\kb,\e,\kb',\e'}
	\left[ 
	\begin{aligned}
	&\phicl^{i}(\kb-\kb', \e - \e') 
	\bar{\psi}^{i}_{\sigma}(\kb,\e) 
	\mf(\e)	 \mf(\e') 
	\psi^{i}_{\sigma}(\kb',\e') 
	\\
	+
	&\phiq^{i}(\kb-\kb', \e - \e')
	\bar{\psi}^{i}_{\sigma}(\kb,\e) 
	\mf(\e) \htau^1 \mf(\e')
	\psi^{i}_{\sigma}(\kb',\e')
	\end{aligned}
	\right].
	\end{aligned}
	\end{align}
	\end{subequations}
\end{widetext}

As mentioned in Sec.~\ref{sec:Kubo}, the drag viscosity and drag resistivity are given by the retarded correlation functions of interacting stress tensor and drag force, respectively. For simplicity, from now on we use a generalized force $\tF$ to indicate both the zero momentum Fourier component of the interacting stress tensor $T_{xy}^{(\msf{int})}(\vex{0})$ and the drag force $\mp i F^{\LL/\RR}_x$. 
Prior to the transformation Eq.~\ref{eq:CoV}, $\tF$ assumes the following form on branch $a$ of the contour
\begin{align}\label{eq:tF}
\begin{aligned}
	\tF_{a}(\WW)
	=\,&
	\intl{\qb,\ww}\intl{\pb,\e}\intl{\pb',\e'}
	f(\vex{q})
	\left[ 
	\bar{\psi}^{\LL}_{a,\sigma}  (\pb',\e')	
	\psi^{\LL}_{a,\sigma}  (\pb'+\qb,\e'+\ww+\WW)
	\right] 
	\\
	&\times 
	\left[ 
	\bar{\psi}^{\RR}_{a,\sigma} (\pb+\qb,\e+\ww) 
	\psi^{\RR}_{a,\sigma}  (\pb,\e)	
	\right],
\end{aligned}
\end{align}
where $f(\vex{q})$ is given by 
\begin{align}\label{eq:fq}
	f(\vex{q})
	\equiv \,&
	\begin{cases}
	\dfrac{q_x q_y}{2q} V'(q),
	&
	\tF(\WW) = \Txy(\vex{0},\WW),	
	\\
	\\
	q_x V_{\msf{inter}}(q),
	&
	\tF(\WW) = \mp i F^{L/R}_x(\WW).
	\end{cases}
\end{align}

The calculation of the drag viscosity and drag resistivity reduces to the that of
the retarded correlation function of $\tF$ :
\begin{align}\label{eq:GF0}
\GG^{(R)}_{\tF}(\WW)=-i\braket{\tF_{\msf{cl}}(\WW)\tF_{\msf{q}}(-\WW)}.
\end{align} 
Here $\tF_{\msf{cl}}$ and $\tF_{\msf{q}}$ stand for, respectively, $\tF$'s classical and quantum components, 
\begin{align}
\tF_{\msf{cl}} \equiv \left( \tF_{1}+\tF_{2}\right) /\sqrt{2},
\qquad
\tF_{\msf{q}} \equiv \left( \tF_{1}-\tF_{2}\right) /\sqrt{2}.
\end{align}
After the transformation Eq.~\ref{eq:CoV}, they acquire the form 
\begin{widetext}
\begin{align}\label{eq:tF2}
\begin{aligned}
	&\tF_{\msf{cl}/\msf{q}}(\WW)
	=\,
	\frac{1}{2\sqrt{2}}
	\intl{\pb,\pb',\e,\e',\qb,\ww}
	f(\vex{q})
	\\
	\times&
	\left\lbrace 
	\left[ 
	\bar{\psi}^{L}_{\sigma'}  (\pb',\e')	
	\mf(\e') 
	\htau^1
	\mf(\e'+\ww+\WW)
	\psi^{L}_{\sigma'}  (\pb'+\qb,\e'+\ww+\WW)
	\right] 
	\left[ 
	\bar{\psi}^{R}_{\sigma} (\pb+\qb,\e+\ww) 
	\mf(\e+\ww) 
	M_1
	\mf(\e)
	\psi^{R}_{\sigma}  (\pb,\e)	
	\right] 
	\right. 
	\\
	&+
	\left. 
	\left[ 
	\bar{\psi}^{L}_{\sigma'}  (\pb',\e')
	\mf(\e')	
	\mf(\e'+\ww+\WW) 
	\psi^{L}_{\sigma'}  (\pb'+\qb,\e'+\ww+\WW)
	\right] 
	\left[ 
	\bar{\psi}^{R}_{\sigma} (\pb+\qb,\e+\ww) 
	\mf(\e+\ww) 
	M_2
	\mf(\e) 
	\psi^{R}_{\sigma}  (\pb,\e)	
	\right] 
	\right\rbrace.
\end{aligned}
\end{align}
\end{widetext}
Here, $M_{1}$ and $M_{2}$ are matrices acting only on the Keldysh space, and are given by: $M_1=\htau^{1}$ ($M_1=1$) and $M_2=1$ ($M_2=\htau^1$) for the classical (quantum) component $\tF_{\msf{cl}}$ ($\tF_{\msf{q}}$).

\subsection{Random Phase Approximation}

\begin{figure}[b!]
	\centering
	\includegraphics[width=0.95\linewidth]{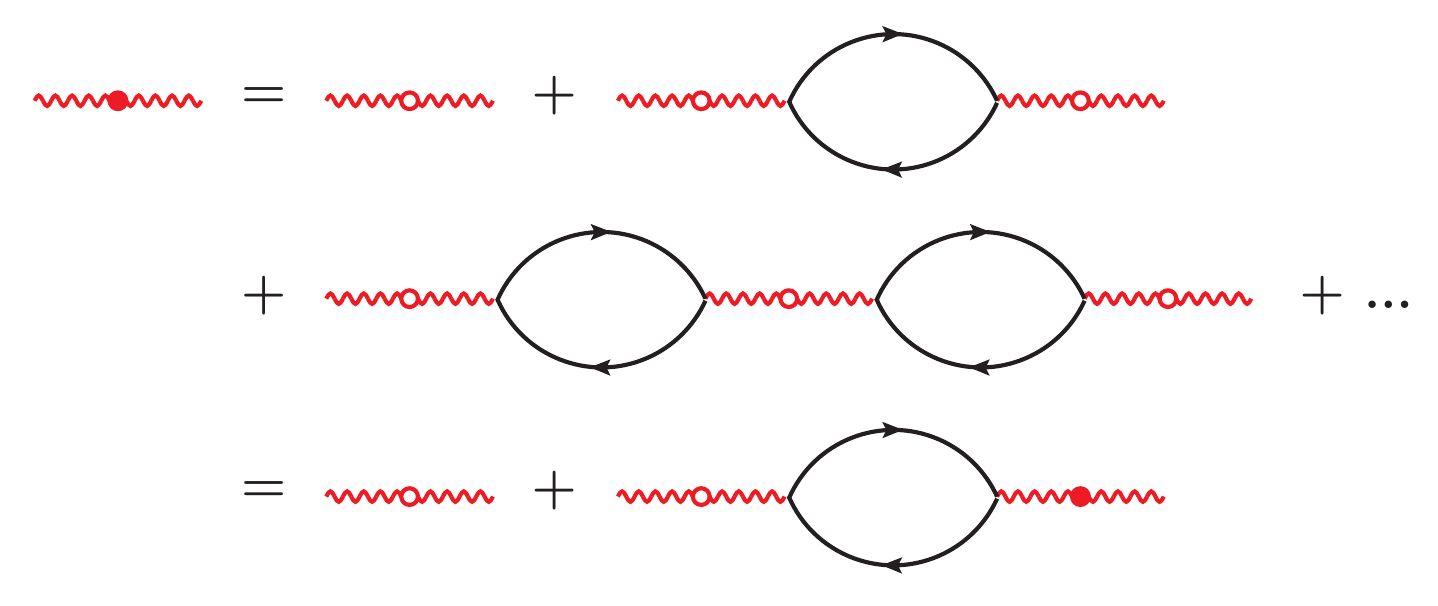}
	\caption
	{
		The Dyson equation for the dressed Green's function $D$ of the H.S. field $\phi$.
		The red wavy line with a solid dot (open
		circle) in the middle refers to the dressed (bare) propagator of the H.S. field $\phi$. 
		The black line with arrow corresponds to the bare propagator $G_0$ for the fermionic field $\psi$.
	}
	\label{fig:D}
\end{figure}

\begin{figure}[b!]
	\centering
	\includegraphics[width=0.90\linewidth]{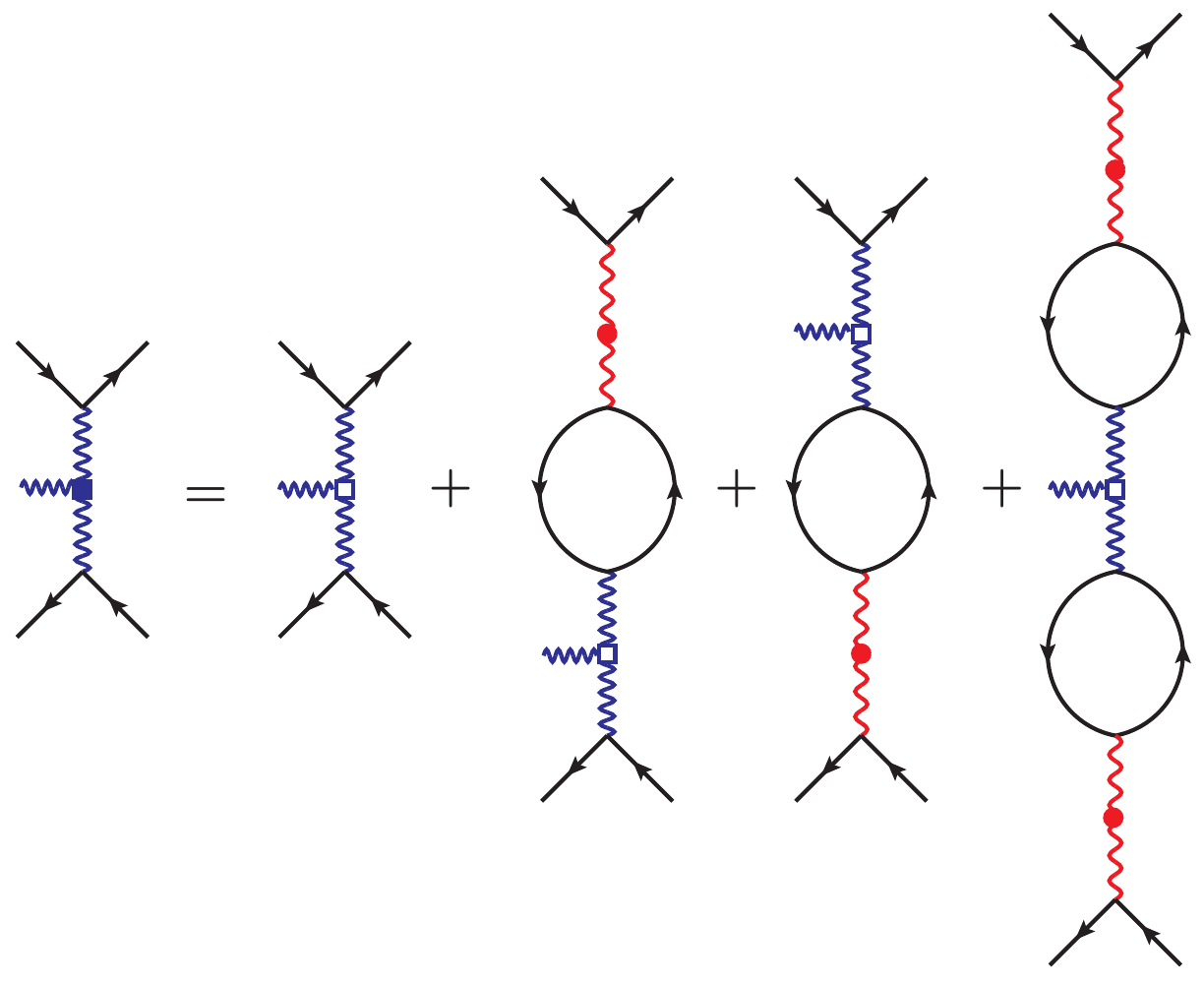}
	\caption{The diagrammatic definition of the dressed interacting stress tensor and dressed drag force. 
		The dressed force $\tF^{(d)}$ is defined as the summation of all the diagrams on the right-hand side of the equal sign, and the  bare force $\tF$ is described by the second diagram in this figure.
		The dressed force $\tF^{(d)}$ and bare force $\tF$ are differentiated by the solid and open blue box in the middle.
		Note that the blue wavy line with a box is used to characterize the vertex $f(\qb)$ of the force $\tF$.
		The layer and spin indices carried by the fermionic fields indicated by the black lines here are identical on the same side of the blue wavy line.
	}
	\label{fig:T}
\end{figure}

It is useful to obtain first the dressed propagator of the H.S. field $\phi$. Integrating out the fermionic field $\psi$ in Eq.~\ref{eq:Z-rot}, we arrive at an effective action $iS_{\phi}+\braket{\left( iS_c \right)^2/2}$ to the leading order in cumulant expansion.
Here $\braket{\left( iS_c \right)^2/2}$ denotes the expectation value of $\left( iS_c \right)^2/2$ (Eq.~\ref{eq:Sc}) with respect to the weight $\exp \left( iS_{\psi}\right)$ (Eq.~\ref{eq:Spsi}),  and can be expressed as
\begin{align}\label{eq:Seff}
\begin{aligned}
	\braket{\frac{1}{2}(iS_c)^2}
	\equiv\,&
	\int \dd \left( \bpsi, \psi \right)
	e^{iS_{\psi}}  \frac{1}{2}(iS_c)^2
	\\
	=\,&
	-\frac{i}{2} 
	\intl{\qb,\ww}
	\phi(-\qb,-\ww) 
	\Pi (\qb,\ww)
	\phi(\qb,\ww). 
\end{aligned}
\end{align}
$\Pi$ is the self energy of the H.S. field $\phi$ and possesses the following causality structure in the Keldysh space
\begin{align}\label{eq:Pi1}
\begin{aligned}
	&\Pi(\qb,\ww)
	=\,
	\begin{bmatrix}
	0& \Pi^{(A)}(\qb,\ww)
	\\
	\Pi^{(R)}(\qb,\ww)& \Pi^{(K)}(\qb,\ww)
	\end{bmatrix}.
\end{aligned}
\end{align}
Its retarded component $\Pi^{(R)}(\qb,\ww)$ is given by
\begin{align}\label{eq:Pi2}
\begin{aligned}
	\Pi^{(R)}_{ij}(\qb,\ww)
	=\,&
	-i \delta_{ij}
	\intl{\kb,\e}
	\left[ 
	G_0^{(R)}(\kb+\qb,\e+\ww)G_0^{(K)}(\kb,\e)
	\right. 
	\\
	&+
	\left. 
	G_0^{(K)}(\kb+\qb,\e+\ww)G_0^{(A)}(\kb,\e)
	\right].
\end{aligned}
\end{align}
$G_0^{(K)}$ indicates the non-interacting Keldysh Green's function of the fermionic field $\psi$, and is related to its retarded and advanced counterparts (Eq.~\ref{eq:G0}) through the fluctuation-dissipation theorem (FDT): 
\begin{align}
	G_0^{(K)}(\kb,\e)=\left[G_0^{(R)}(\kb,\e)-G_0^{(A)}(\kb,\e) \right]\tanh \left( \ww/2T\right).
\end{align}
From Eq.~\ref{eq:Seff}, one can also prove that self energy's advanced $\Pi^{(A)}$ and Keldysh $\Pi^{(K)}$ components are related to the retarded one $\Pi^{(R)}$ through
\begin{align}\label{eq:Pi3}
\begin{aligned}
	\Pi^{(A)}(\qb,\ww)
	=\,&
	\left[ \Pi^{(R)}(\qb,\ww)\right]^{\dagger} ,
	\\
	\Pi^{(K)}(\qb,\ww)=\,&
	\left[  \Pi^{(R)}(\qb,\ww) -\Pi^{(A)} (\qb,\ww)\right]
	 \coth \left(\ww/2T\right).
\end{aligned}
\end{align}
The last equation is a manifestation of the FDT relation between the components of bosonic self energy.

The dressed propagator for $\phi$ can be deduced from the effective quadratic action $iS_{\phi}+\braket{\left( iS_c \right)^2/2}$:
\begin{align}\label{eq:D1}
\begin{aligned}
	D_{ab;ij}(\qb,\ww)
	\equiv &
	-i
	\braket{
		\phi_a^{i}(\qb,\ww)
		\phi_b^{j}(-\qb,-\ww)
	}
	\\
	=\,&
	\left[ 
	D_0^{-1}(\qb,\ww)
	-
	\Pi(\qb,\ww)
	\right]^{-1}_{ab;ij}.
\end{aligned}
\end{align}
$D_0$ here represents the bare propagator for the H.S. field $\phi$ and takes the form
\begin{align}\label{eq:D0}
	D_0(\qb,\ww)
	=
	\begin{bmatrix}
	0 & V(q)
	\\
	V(q) & 0
	\end{bmatrix},
\end{align}
where the interaction potential $V(q)$ is given by Eq.~\ref{eq:V} in the layer space.
Inserting Eq.~\ref{eq:Pi1} into the Dyson Eq.~\ref{eq:D1}, one finds the $D(\qb,\ww)$ acquires the typical causality structure of a bosonic field's propagator:
\begin{align}\label{eq:D2}
\begin{aligned}	
	D(\qb,\ww)=\,&
	\begin{bmatrix}
	D^{(K)}(\qb,\ww)& D^{(R)}(\qb,\ww)
	\\
	D^{(A)}(\qb,\ww)& 0
	\end{bmatrix},
\end{aligned}
\end{align}
and its retarded $D^{(R)}$, advanced $D^{(A)}$ and Keldysh $D^{(K)}$ components are given by, respectively
\begin{align}\label{eq:D3}
\begin{aligned}
	&D^{(R)}(\qb,\ww)
	=\,
	\left[ V^{-1}(q)-\Pi^{(R)}(\qb,\ww)\right]^{-1},
	\\
	&D^{(A)}(\qb,\ww)
	=\,
	\left[ D^{(R)}(\qb,\ww)\right]^{\dagger} ,
	\\
	&D^{(K)}(\qb,\ww)
	=\,
	\left[ D^{(R)}(\qb,\ww)-D^{(A)}(\qb,\ww)\right] \coth \left(\ww/2T\right). 
\end{aligned}
\end{align}
As indicated by the last equation, the dressed $\phi$ propagator $D(\qb,\ww)$ satisfies the FDT relation.

We note that the leading order cumulant expansion performed above is equivalent to the random phase approximation (RPA) valid in the high density ($\rs \ll 1$) limit.
The dressed propagator $D(\qb,\ww)$ given by Eqs.~\ref{eq:D2} and~\ref{eq:D3} is the RPA interaction, and is represented diagrammatically by the Dyson series in Fig.~\ref{fig:D}. Hereafter, we use the red wavy line with a solid dot (open
circle) in the middle to describe the dressed (bare) propagator of the H.S. field $\phi$. 
The bosonic self energy  (or the polarization operator) $\Pi$ is given by the black bubble in Fig.~\ref{fig:D} with black line representing the bare propagator of the fermionic field $\psi$.

\subsection{Retarded Correlation Function of Stress Tensor and Drag Force}\label{sec:Cor}

We now calculate the retarded correlation function $\GG^{(R)}_{\tF}(\WW)$ of the force $\tF$ (Eq.~\ref{eq:GF0}).
The diagrammatic representation of the bare force $\tF$ is given by the second diagram in Fig.~\ref{fig:T}, where the blue wavy line with a open box in the middle corresponds to the vertex $f(\qb)$ (Eq.~\ref{eq:fq}). On the same (opposite) side of this line, fermionic fields denoted by black lines carry identical (opposite) layer and spin indices.
 
First, we introduce a source field $J $ which couples to the bilinear combination of the fermionic field $\bpsi \psi$ in the same way as the bosonic field $\phi$. More specifically, we apply the shift $\phi \rightarrow \phi +J $ in the part of the action $S_c$ (Eq.~\ref{eq:Sc}) which describes the coupling between $\psi$ and $\phi$, and leave the other two parts $S_{\psi}$ (Eq.~\ref{eq:Spsi}) and $S_{\phi}$ (Eq.~\ref{eq:Sphi}) unchanged:
\begin{widetext}
\begin{align}
\begin{aligned}
	Z[J]
	=
	\int 
	\D \left( \bar{\psi},\psi \right) 
	\D \phi
	\exp
	&
	\left\lbrace 
	iS_{\phi}[\phi]
	+
	iS_{\psi}[\bpsi,\psi]
	+
	iS_{c} [\phi+J,\bpsi, \psi]
	\right\rbrace.
\end{aligned}
\end{align}

Integrating out the fermionic degrees of freedom, one arrives at
\begin{align}
\begin{aligned}
	&Z[J]
	=
	\int 
	\D \phi
	\exp
	\left\lbrace 
	\frac{i}{2} 
	\intl{\qb,\ww}
	\phi(-\qb,-\ww) 
	D_0^{-1}(\qb,\ww)
	\phi(\qb,\ww) 
	-\frac{i}{2} 
	\intl{\qb,\ww}
	\left[ \phi(-\qb,-\ww)+J(-\qb,-\ww)\right] 
	\Pi(\qb,\ww)
	\left[  \phi(\qb,\ww)+J(\qb,\ww) \right] 
	\right\rbrace. 
\end{aligned}
\end{align}
The retarded correlation function $\GG^{(R)}_{\tF}(\WW)$ can be obtained by taking the functional derivatives of $Z[J]$ with respect to the source field $J$ and then setting $J$ to zero,
\begin{align}\label{eq:GF-1}
\begin{aligned}
	i\GG^{(R)}_{\tF}(\WW)
	=\,&
	\frac{1}{2}
	\intl{\ww,\qb}\intl{\ww',\qb'}
	f(\vex{q})f(-\vex{q}')
	\left\lbrace 
	\dfrac{\delta^2}
	{\delta \Jq^{L}(-\qb, -\ww-\Omega)\delta \Jq^{R}(\qb, \ww)}
	+
	\dfrac{\delta^2}
	{\delta \Jcl^{L}(-\qb, -\ww-\Omega)\delta \Jcl^{R}(\qb, \ww)}
	\right\rbrace 
	\\
	\times &
	\left\lbrace 
	\dfrac{\delta^2}{\delta \Jq^{L}(\qb', \ww'+\WW) \delta \Jcl^{R}(-\qb', -\ww')}
	+
	\dfrac{\delta^2}{\delta \Jcl^{L}(\qb', \ww'+\WW)\delta \Jq^{R}(-\qb', -\ww')}
	\right\rbrace 
	Z[J]|_{J=0}.
\end{aligned}
\end{align} 
Here we have utilized the fact that $Z[J=0]=1$.
Using the RPA approximation, we argue that only the contributions from $\qb'=\qb,\ww'=\ww$ and $\qb'=-\qb,\ww'=-\ww-\WW$ to the integral in Eq.~\ref{eq:GF-1} are important and will be considered in what follows.

Each term in Eq.~\ref{eq:GF-1} (with $\qb'=\qb,\ww'=\ww$ or $\qb'=-\qb,\ww'=-\ww-\WW$) is of the form
\begin{align}\label{eq:dJ}
\begin{aligned}
 	&
 	\dfrac{\delta^4Z[J]}{\delta J_a^{i}(-\qb, -\ww-\WW)\delta J_b^{j}(\qb, \ww+\WW)\delta J_c^{m}(\qb, \ww)\delta J_d^{n}(-\qb, -\ww)}\biggr\rvert_{J=0}
 	\\
 	&=
 	-
 	\left[ 
 	\Pi(\qb,\ww+\WW)
 	+
 	 \Pi(\qb,\ww+\WW) D (\qb,\ww+\WW)\Pi(\qb,\ww+\WW)
 	 \right]^{ab}_{ij}
	\left[ 
 	\Pi(-\qb,-\ww)
 	+
 	 \Pi(-\qb,-\ww) D (-\qb,-\ww)\Pi (-\qb,-\ww)
 	 \right]^{cd}_{mn}
 	 \\
 	 &=
 	 -
 	  \left[ 
 	  D_0^{-1}(\qb,\ww+\WW)
 	  D (\qb,\ww+\WW)
 	 \Pi(\qb,\ww+\WW) 
 	 \right]^{ab}_{ij}
 	 \left[ 
 	 D_0^{-1}(-\qb,-\ww)
 	 D (-\qb,-\ww)
 	 \Pi(-\qb,-\ww) 
 	 \right]_{mn}^{cd},
\end{aligned}	
\end{align}
where in the last equality, the Dyson Eq.~\ref{eq:D1} has been employed.

Combining the equation above with Eqs.~\ref{eq:D0},~\ref{eq:Pi1} and~\ref{eq:D2}, we find $i\GG_{\tF}^{(R)}$ reduces to
	\begin{align}\label{eq:GF-2}
	\begin{aligned}
	&i\GG^{(R)}_{\tF}(\WW)=\,
	-\frac{a_{\rm L}}{2}
	\suml{i\in \left\lbrace \LL,\RR\right\rbrace }
	\int_{\ww,\qb}
	f(\vex{q})f(-\zeta_i \vex{q})
	\\
	&
	\left\lbrace 
	\begin{aligned}
	&
	\left[ 
	V^{-1}(q)
	\left( 
	D^{(K)}(\qb,\ww+\WW)
	\Pi^{(A)}(\qb,\ww+\WW)
	+
	D^{(R)}(\qb,\ww+\WW)
	\Pi^{(K)}(\qb,\ww+\WW)
	\right) 
	\right]_{\LL i} 
	\left[ 
	V^{-1}(q)
	D^{(R)}(-\qb,-\ww)
	\Pi^{(R)}(-\qb,-\ww)
	\right]_{\RR \bar{i}} 
	\\
	&
	+
	\left[ 
	V^{-1}(q)
	D^{(R)}(\qb,\ww+\WW)
	\Pi^{(R)}(\qb,\ww+\WW)
	\right]_{\LL i} 
	\left[ 
	V^{-1}(q)
	\left( 
	D^{(K)}(-\qb,-\ww)
	\Pi^{(A)}(-\qb,-\ww)
	+
	D^{(R)}(-\qb,-\ww)
	\Pi^{(K)}(-\qb,-\ww)
	\right) 
	\right]_{\RR \bar{i}} 
	\end{aligned}
	\right\rbrace, 
	\end{aligned}
	\end{align}
whose imaginary part takes the form
\begin{align}\label{eq:ImGF}
\begin{aligned}
&\im \GG^{(R)}_{\tF}(\WW)=
\frac{a_{\rm L}}{4}
\suml{i\in\left\lbrace \LL,\RR\right\rbrace}
\int_{\ww,\qb}
f(\vex{q})f(-\zeta_i \vex{q})
\left[ \coth \left( \frac{\ww+\WW}{2T} \right) - \coth \left( \frac{\ww}{2T} \right)\right] 
\\
\times&
\left\lbrace 
\left[ 
V^{-1}(q)
\left( 
D^{(R)}(\qb,\ww+\WW)
\Pi^{(R)}(\qb,\ww+\WW)
-
D^{(A)}(\qb,\ww+\WW)
\Pi^{(A)}(\qb,\ww+\WW)
\right) 
\right]_{\LL i} 
\right. 
\\
&\times
\left. 
\left[ 
V^{-1}(q)
\left( 
D^{(R)}(-\qb,-\ww)
\Pi^{(R)}(-\qb,-\ww)
-
D^{(A)}(-\qb,-\ww)
\Pi^{(A)}(-\qb,-\ww)
\right) 
\right]_{\RR \bar{i}}  
\right\rbrace. 
\end{aligned}
\end{align}
Here the sign factor $\zeta_i=1$ $(-1)$ when $i=\LL$ $(\RR)$, and $\bar{i}$ is defined such that $\bar{\LL}=\RR$ and $\bar{\RR}=\LL$. 
The combinatorial factor $a_{\rm L}$ takes the value of 2 (1), for $\tF = \Txy(\vex{0})$ ($\mp i F^{\LL/\RR}_x$).
This factor accounts for the contributions from $\qb'=\qb,\ww'=\ww$ and $\qb'=-\qb,\ww'=-\ww-\WW$ to the retarded correlation function of the interacting stress tensor (given by the integral in Eq.~\ref{eq:GF-1}).
\end{widetext}

The retarded correlation function $\GG_{\tF}^{(R)}$ in Eq.~\ref{eq:GF-2} is represented by the diagrams in Fig.~\ref{fig:E}, which can also be considered as the non-interacting correlation function of the dressed force $\tF^{(d)}$ and the bare force $\tF$ (see Eq.~\ref{eq:GF-a1} in Appendix~\ref{app:diagram}).
The dressed force $\tF^{(d)}$ is defined diagrammatically in Fig.~\ref{fig:T}, and its explicit expression is given by Eq.~\ref{eq:dF}.
The dressed and bare forces, illustrated respectively as the first and second diagrams in Fig.~\ref{fig:E}, can be distinguished by the box in the middle: the solid (open) box corresponds to the dressed (bare) force. 
From Fig.~\ref{fig:T}, i.e., the diagrammatic definition of the dressed force $\tF^{(d)}$, one can immediately see that the non-interacting correlation function of the dressed and bare forces (see the first diagram in Fig.~\ref{fig:E}) is equivalent to the summation of diagrams shown on the right-hand side of the equal sign in Fig.~\ref{fig:E}.
Note that the red wavy line with a red solid dot in the middle is the dressed propagator of the H.S. field $\phi$ which is given by the infinite geometric series in Fig.~\ref{fig:D}.
As a result, $\GG_{\tF}^{(R)}$ is also an infinite diagrammatic series with repeated insertion of polarization bubbles.
See Appendix~\ref{app:diagram} for a diagrammatic derivation of $\GG_{\tF}^{(R)}$ (Eq.~\ref{eq:ImGF}).

\subsection{General Formulas for Drag Viscosity and Drag Resistivity}

Substituting Eqs.~\ref{eq:ImGF} and~\ref{eq:fq} into the Kubo formulas Eqs.~\ref{eq:etaD} and~\ref{eq:Kubo-R}, and expressing the dressed propagator $D$ in terms of the self energy $\Pi$ and the bare propagator $D_0$ (Eq.~\ref{eq:D3}), we find the general expressions for the drag viscosity 
\begin{widetext}
\begin{align}
	\label{eq:drag-eta}
	&\eta_{\msf{D}}
	=\,
	\intl{\qb,\ww}
	\dfrac{	\left[ \dfrac{q_x q_y}{2q} V'(q) V^{-2}(q) \right]^2}{T\sinh^2 \left( \frac{\ww}{2T}\right) }
	\dfrac
	{
		\left[ 
		\im
		\Pi^{(R)}(\qb,\ww)
		\right]^2
	}
	{
		\left\lbrace 
		\left[ 
		V^{-1}(q) 
		-
		\re \Pi^{(R)}(\qb,\ww)
		\right] ^2
		+
		\left[ 
		\im \Pi^{(R)}(\qb,\ww)
		\right] ^2 
		\right\rbrace^2
		},
\end{align}
and for the drag resistivity
\begin{align}
	\label{eq:drag-rho}
	&\begin{aligned}
	\rho_{\msf{D}}
	=\, &
	\frac{1}{2n_{\LL} n_{\RR} e^2}
	\int_{\ww,\qb}
	\dfrac{\left[ q_x V_{\msf{inter}}(q) \right]^2}{T\sinh^2 \left( \frac{\ww}{2T}\right)}
	\\
	\times&
	\dfrac{	 \im \Pi^{(R)}_{\LL\LL}(\qb,\ww) \im \Pi^{(R)}_{\RR\RR}(\qb,\ww)  }{
	\left| 
	-1
	+
	\left[ 
	\Pi^{(R)}_{\LL\LL}(\qb,\ww)
	+
	\Pi^{(R)}_{\RR\RR}(\qb,\ww)
	\right] 
	V_{\msf{intra}}(q)
	+
	\Pi^{(R)}_{\LL\LL}(\qb,\ww)
	\Pi^{(R)}_{\RR\RR}(\qb,\ww)
	\left( 
	V_{\msf{inter}}^2(q)-V_{\msf{intra}}^2(q)
	\right) 	
	\right|^2 
	}.
	\end{aligned}
\end{align}
Here we have used the fact that the bosonic self energy $\Pi^{(R)}$ is diagonal in the layer space.
In the case where the two layers are identical, $\Pi^{(R)}_{\LL\LL}=\Pi^{(R)}_{\RR\RR}$, the expression for drag resistivity $\rho_D$ reduces to
\begin{align}\label{eq:drag-R-1}
\begin{aligned}
	\rho_{\msf{D}}
	=\,&
	\frac{1}{2n^2 e^2}
	\int_{\ww,\qb}
	\dfrac{	\left[ q_x V_{\msf{inter}}(q) \right]^2}{T\sinh^2 \left( \frac{\ww}{2T}\right) }
	\dfrac{	\left[ \im \Pi^{(R)}_{\LL \LL} (\qb,\ww)\right]^2 }{
		\left| 
		1+
		\Pi^{(R)}_{\LL \LL}(\qb,\ww)
		\left( 
		V_{\msf{inter}}(q)-V_{\msf{intra}}(q)
		\right) 	
		\right|^2 
		\left| 
		1-
		\Pi^{(R)}_{\LL \LL}(\qb,\ww)
		\left( 
		V_{\msf{inter}}(q)+V_{\msf{intra}}(q)
		\right) 	
		\right|^2 
	}.
\end{aligned}
\end{align}
Hereafter, for simplicity, we consider only the case of two identical layers 
and use $\Pi^{(R)}(\qb,\ww)$ to indicate the single layer retarded polarization operator, which is previously denoted as $\Pi^{(R)}_{\LL \LL/\RR \RR}(\qb,\ww)$.
\end{widetext}	

Eq.~\ref{eq:drag-R-1} was derived earlier by Chen et al.~\cite{Langevin} through the Boltzmann-Langevin approach, whereas Eq.~\ref{eq:drag-eta} is obtained for the first time for the drag contribution to viscosity.
We want to emphasize the close resemblance between these two expressions.
Nowhere in the derivation here have we used the specific forms of interaction potential $V(q)$ and that of the dispersion $\xi_{\kb}$. 
The general formulas for drag viscosity and drag resistivity (Eqs.~\ref{eq:drag-eta} and~\ref{eq:drag-R-1}), are therefore applicable to other interacting fermionic systems described by Eq.~\ref{eq:ZK}. 
Long-range interactions with any form of bare potential $V(q)$ will induce frictional drag forces acting on the charge carries, which result in drag viscosity and Coulomb drag. We note that the bare potential $V(q)$ enters the formula for Coulomb drag while the derivative $V’(q)$ enters that for drag viscosity, which is strong only with large $V’(q)$.
Furthermore, RPA approximation employed here is only valid in the high density limit ($\rs \ll 1$). To go beyond RPA, one way to estimate the result is by replacing the noninteracting polarization operator $\Pi^{(R)}$ with the interacting one~\cite{Maslov}, as in Ref.~\cite{Langevin} for Coulomb drag.
We note that the expression for the retarded correlation function $\GG_{\tF}^{(R)}$ can also be used to deduce the dynamical drag viscosity~\cite{Conti} using $\eta_{\msf{D}}(\ww)=-\im \GG_{T}^{(R)}(\vex{0},\ww)/\ww$.

\begin{figure}[t!]
	\centering
	\includegraphics[width=0.99\linewidth]{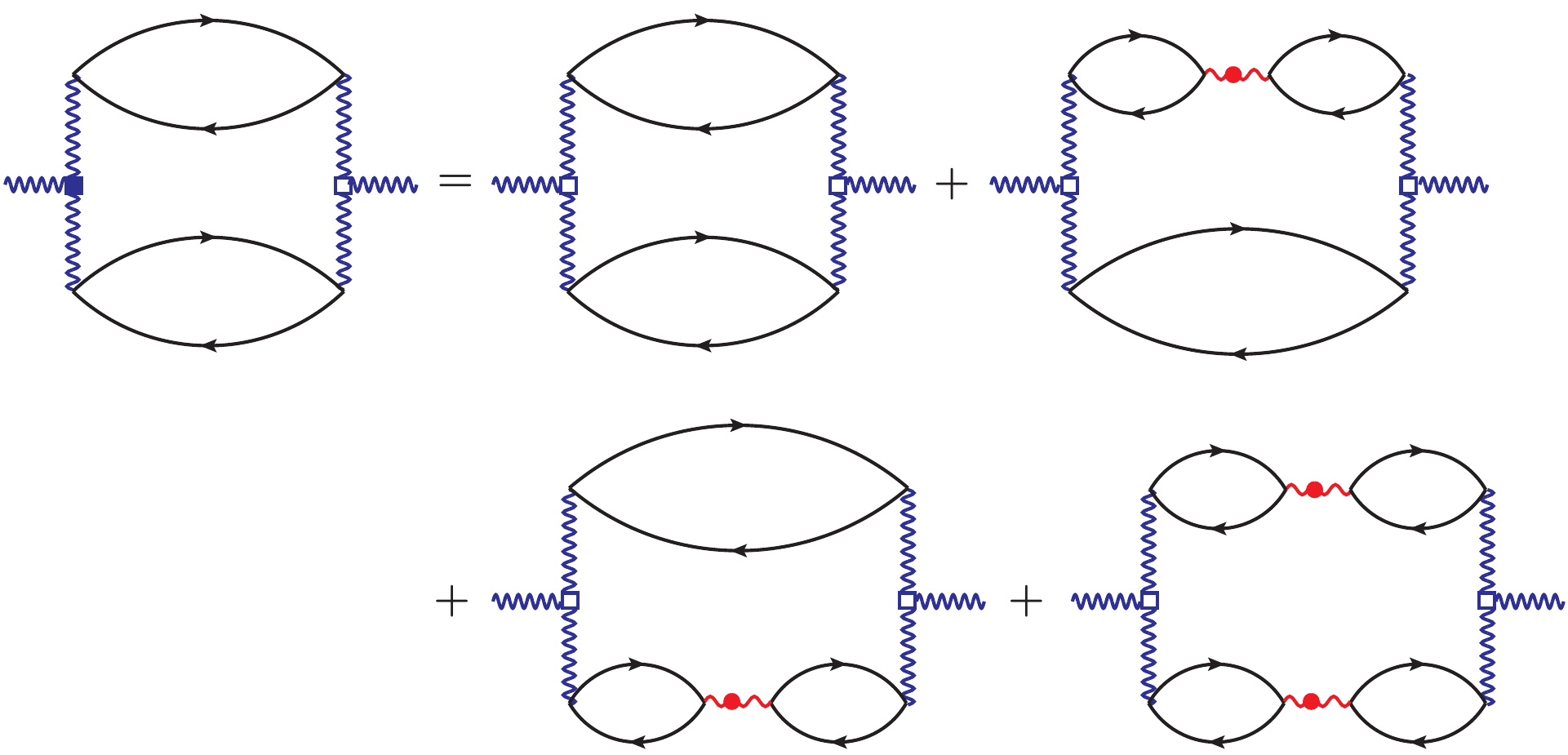}
	\caption{The retarded correlation function $\GG_{\tF}^{(R)}$ of the force $\tF$ is given by the non-interacting correlation function of the bare force $\tF$ and the dressed one $\tF^{(d)}$ which is defined diagrammatically in Fig.~\ref{fig:T}. It is equivalent to the summation of the four diagrams shown on the right-hand side of the equal sign in this figure.}
	\label{fig:E}
\end{figure}

The general formulas for the drag viscosity $\eta_{\msf{D}}$ (Eq.~\ref{eq:drag-eta})
and drag resistivity $2n^2e^2\rho_{\msf{D}}$ (Eq.~\ref{eq:drag-R-1}) can be rewritten as 
\begin{align}\label{eq:general}
\begin{aligned}
	\int_{\qb,\ww} \tilde{f}^2(\qb,\ww) R(\qb,\ww),
\end{aligned}	
\end{align} 
where $\tilde{f}(\qb,\ww)$ acquires the form
\begin{align}\label{eq:tfq}
\tilde{f}(\vex{q},\ww)
\equiv \,&
\begin{cases}
f(\qb)
\left\lvert\dfrac{ D^{(R)}(\qb,\ww) }{V(q)}\right\rvert^2,
&
\text{for} \, \eta_{\msf{D}},
\\
\\
f(\qb)
\left\lvert\dfrac{ D^{(R)}_{\LL \RR}(\qb,\ww) }{V_{\msf{inter}}(q)} \right\rvert,
&
\text{for}\, \rho_{\msf{D}},
\end{cases}
\end{align}
 and
$R(\qb,\ww)$ is defined as
\begin{align}\label{eq:R}
	R(\qb,\ww)
	\equiv\,
	\dfrac{\left[ \im \Pi^{R} (\qb,\ww)\right]^2 }{ T\sinh^2 \left( \frac{\ww}{2T} \right) }.
\end{align}
If one ignores the interaction correction to $\GG_{\tF}^{(R)}$ (i.e., considering only the second diagram in Fig.~\ref{fig:E} and neglecting the contribution from the rest), the resulting drag viscosity and drag resistivity can also be expressed as Eq.~\ref{eq:general} with $\tilde{f}(\qb,\ww)$ replaced by the bare vertex $f(\qb)$ (Eq.~\ref{eq:fq}).

Insertion of the single-particle fermionic Green's function $G_0$ (Eq.~\ref{eq:G0}) into Eq.~\ref{eq:Pi2} leads to 
\begin{align}\label{eq:Pi4}
\begin{aligned}
	\Pi^{(R)} (\qb,\ww)
	=\,&
	\intl{\kb}
	\dfrac{\tanh\left( \frac{\xi_{\kb+\qb}}{2T} \right) 
		-\tanh\left(\frac{\xi_{\kb}}{2T} \right)}
	{\ww+\xi_{\kb}- \xi_{\kb+\qb}+i\eta },
\end{aligned}
\end{align}
whose imaginary part can be expressed as
\begin{align}\label{eq:ImPi}
\begin{aligned}
	&\im
	\Pi^{(R)}(\qb,\ww)
	\\
	=&
	2\pi 
	\intl{\kb}
	\delta \left( \ww-\xi_{\kb+\qb} +\xi_{\kb} \right) 
	n_{\msf{F}}(\xi_{\kb})
	\left[1-n_{\msf{F}}(\xi_{\kb+\qb}) \right] 
	\left(e^ {-\frac{\ww}{T} }  -1 \right) 
	\\
	=\,&
	2\pi 
	\intl{\kb}
	\delta \left( \ww-\xi_{\kb+\qb} +\xi_{\kb} \right) 
	\left[1-n_{\msf{F}}(\xi_{\kb})\right] 
	n_{\msf{F}}(\xi_{\kb+\qb})
	\left(  1- e^ {\frac{\ww}{T} } \right).
\end{aligned}
\end{align}
Here $n_{\msf{F}}(\xi)=1/\left(e^{\xi/T}+1\right) $ is the Fermi distribution function.
Making use of Eq.~\ref{eq:ImPi}, one finds that $R(\qb,\ww)$ can be written as~\cite{Rojo}
\begin{align}\label{eq:drag-eta-s2}
\begin{aligned}
	&R(\qb,\ww)
	=\,
	\dfrac{\left[ \im \Pi^{R} (\qb,\ww)\right]^2 }{ T\sinh^2 \left( \frac{\ww}{2T} \right) }
	\\
	=\,&
	\frac{16\pi^2}{T}
	\intl{\kb_1,\kb_2}
	\delta \left(\ww+\xi_{\kb_1} -\xi_{\kb_1+\qb} \right) 
	\delta \left(\ww-\xi_{\kb_2} +\xi_{\kb_2-\qb}\right) 
	\\
	&\times
	n_{\msf{F}}(\xi_{\kb_1})
	\left[1-n_{\msf{F}}(\xi_{\kb_1+\qb}) \right] 
	n_{\msf{F}}(\xi_{\kb_2})
	\left[1-n_{\msf{F}}(\xi_{\kb_2-\qb}) \right],
\end{aligned}
\end{align}
which is therefore related to the phase space available for interlayer scattering processes with momentum transfer $\qb$ and energy transfer $\ww$.
$R(\qb,\ww)$ is related to the number of particle-hole excitations with momentum $\qb$ and energy $\ww$.
In the present paper, we only consider sufficiently low temperatures at which the contribution from particle-hole continuum is dominant and that from the plasmons can be ignored. 

\section{Results}~\label{sec:Results}

To calculate the drag viscosity and drag resistivity using Eqs.~\ref{eq:drag-eta} and~\ref{eq:drag-R-1}, we still need the explicit expression for the polarization operator $\Pi^{(R)}$.
Its value at zero temperature was calculated by Lindhard~\cite{lindhard} and Stern~\cite{stern} in 3D and 2D, respectively. In both dimensions, the zero-temperature polarization operator $\Pi^{(R)}(\qb,\ww)$ can be expressed in terms of two dimensionless parameters $u \equiv \ww/\vf q$ and $z=q/2\pf$, with $\pf$, $\vf=\pf/m$ being the Fermi momentum and Fermi velocity, respectively. See Appendix~\ref{app:Pi} for the explicit expressions of $\Pi^{(R)}(\qb,\ww)$. In the following, we only consider sufficiently low temperatures at which the polarization operator can be approximated by its zero-temperature result. Using the finite temperature expression leads to a correction that is of higher order in $T/\Tf$.

\subsection{Drag Viscosity}

Let us first consider the drag viscosity at temperatures $T/\Tf \ll \alpha$. $\alpha \equiv \ktf/\pf$ is the ratio of the Thomas-Fermi screening wavevector $\ktf$ over the Fermi momentum $\pf$.  $\ktf$ takes the value of $\ktf=\sqrt{4\pi e^2 \nu_0}$ ($\ktf=2\pi e^2 \nu_0$) in 3D (2D), with
$\nu_0= m \pf/\pi^2$ $(\nu_0=m/\pi)$ being the density of states at Fermi level. In terms of dimensionless interaction parameter $\rs$, $\alpha=(16/3\pi^2)^{1/3}\sqrt{\rs}$ $(\alpha=\sqrt{2} \rs)$ in 3D (2D).

Because of the screened interactions and the thermal factor $1/\sinh^2 (\ww/2T)$, the most significant contribution to the integral in Eq.~\ref{eq:drag-eta} comes from small momentum $q \lesssim \ktf$ and small frequency $\ww \lesssim T$. As a result, at low temperatures $T/\Tf \ll \alpha$, one may approximate the polarization operator $\Pi^{(R)} (u,z)$ by its value at $u,z \ll1 $, which leads to the following static limit ($\ww \ll \vf q$) expression for drag viscosity
\begin{align}\label{eq:drag-eta-s}
\begin{aligned}
	\eta_{\msf{D}}
	=\,&
	\frac{1}{4}
	\intl{\qb,\ww}
	\left[ \frac{q_x q_y}{q} \tilde{V}'(q) \right]^2
	\dfrac{\left[ 
		\im
		\Pi^{(R)}(\qb,\ww)
		\right]^2}{T\sinh^2 \left( \frac{\ww}{2T}\right) }.
\end{aligned}
\end{align}
Here $\tilde{V}(q)$ denotes the static screened Coulomb potential
\begin{align}
	\tilde{V}(q)
	\equiv 
	\left[ V^{-1}(q)+\nu_0 \right]^{-1} 
	=
	\begin{cases}
	\dfrac{1}{\nu_0}\dfrac{\ktf^2}{q^2+\ktf^2},
	&
	\text{in 3D},
	\\
	\\
	\dfrac{1}{\nu_0}\dfrac{\ktf}{q+\ktf},
	& 
	\text{in 2D}.
	\end{cases}
\end{align}
Eq.~\ref{eq:drag-eta-s} is obtained by replacing the retarded Green's function for the H.S. field $D^{(R)}(\qb,\ww)$ in Eq.~\ref{eq:tfq} with the static screened interaction $\tilde{V}(q)$ and making use of the fact that $V'(q)V^{-2}(q)=\tilde{V}'(q)\tilde{V}^{-2}(q)$.

$\im \Pi^{(R)}(u,z)$ is well approximated by $-\frac{\pi \nu_0}{2} u \Theta (1-|u|)$ $\left[ -\nu_0 u \Theta (1-|u|)\right] $ in 3D (2D) for $u, z\ll 1$.
Inserting these expressions into Eq.~\ref{eq:drag-eta-s}, one obtains the drag viscosity:
\begin{subequations}
\begin{align}
	&\begin{aligned}
	\eta_{\msf{D}}^{(3D)}
	=&
	\frac{1}{240\pi}\frac{\ktf^4}{v_F^2}
	\int_{-\infty}^{\infty} d\ww 
	\dfrac{\ww^2}{T \sinh^2 \left( \frac{\ww}{2T}\right) }
	\int_{\frac{|\ww|}{\vf}}^{\infty}  dq
		\dfrac
	{
			q^4 
	}
	{
		\left( 
		q^2
		+
		\ktf^2
		\right)^4
	}
	\\
	=&
	\frac{\pi^2}{2880} \frac{\ktf T^2}{\vf^2}
	=\,
	\frac{1}{1280}\left( \frac{\pi^7}{12}\right)^{\frac{1}{3}} 
	\frac{1}{\rs^{5/2} a_0^{3} }
	\left( \frac{T}{\Tf} \right)^2, 
	\end{aligned}
	\\
	&
	\begin{aligned}
	\eta_{\msf{D}}^{(2D)}
	= \,&
	 \frac{1}{2^7 \pi^2}\frac{\ktf^2}{\vf^2}
	\int_{-\infty}^{\infty} d\ww 
	\dfrac{\ww^2}{T\sinh^2 \left( \frac{\ww}{2T}\right) }
	\int_{\frac{|\ww|}{\vf}}^{\infty}  dq
	\dfrac
	{
	 q 
	}
	{
		\left( 
		q
		+
		\ktf
		\right)^4
	}
	\\
	=\,&
	\frac{1}{288}\frac{T^2}{\vf^2}
	=\,
	\frac{1}{576} \frac{1}{\rs^{2} a_0^{2}} \left( \frac{T}{\Tf} \right)^2,
	\end{aligned}
\end{align}
\end{subequations}
in 3D and 2D, respectively.
Here in the last equality of both equations, the drag viscosities are expressed in terms of $\rs$, dimensionless temperature $T/\Tf$,  and Bohr radius $a_0=1/m e^2$.

At higher temperatures $T/\Tf \gtrsim \alpha$ which are still small compared to the Fermi temperature $T/\Tf \ll 1$, the dominant contribution arises from $z \ll 1$ and arbitrary $u$ (as long as $\im \Pi^{(R)}(u,z)$ is nonvanishing). For this reason, we use the expression for the polarization operator at $z=0$ and arbitrary $u$, and find 
\begin{subequations}\label{eq:eta}
\begin{align}
	&
	\begin{aligned}\label{eq:eta3d}
	\eta_{\msf{D}}^{(3D)}
	=\,&
	\frac{1}{60\pi} \frac{\ktf^4}{\pf}
	 \frac{\Tf}{T}  
	\int_0^{\infty}  d x \int_{0}^{1} du \, 
	\dfrac{1}{\sinh^2 \left(   \frac{\alpha}{T/\Tf} u x \right)}
	\\
	\times &
	\dfrac
	{
		u^2 x^7
	}
	{
		\left\lbrace 
		\left[
		x^2
		+
		1
		-
		\frac{u}{2}
		\ln 
		\left(   \dfrac{1+u}{1-u} \right) 
		\right]  ^2
		+
		\left( 
		\frac{\pi }{2} u
		\right)^2 
		\right\rbrace^2
			},		
	\end{aligned}
	\\
	&
	\begin{aligned}	\label{eq:eta2d}
	\eta_{\msf{D}}^{(2D)}
	=\,&
	\frac{1}{32\pi^2}\frac{\ktf^3}{\pf}
	\frac{\Tf}{T}
	\int_0^{\infty} d x 
	\int_{0}^{1} d u
	\dfrac{1}{\sinh^2 \left( \frac{\alpha}{T/\Tf} ux\right) }
	\\
	\times &
	\dfrac
	{
		x^4 u^2\left( 1-u^2 \right)
	}
	{
		\left\lbrace 
		\left( 
		x
		+
		1
		\right) ^2
		-
		\left[ 
		\left( 
		x
		+
		1
		\right)^2
		-
		1\right] 
		u^2
		\right\rbrace^2
	},	
	\end{aligned}
\end{align}
\end{subequations}
where $ x \equiv q/\ktf$.

The thermal factor $1/\sinh^2 (\alpha ux \Tf/T)$ in Eq.~\ref{eq:eta} can be approximated by $(T/\Tf \alpha ux  )^2$ for $ T/\Tf \gg \alpha$,  which leads to the linear temperature dependence of $\eta_{\msf{D}}$ in both dimensions:
\begin{subequations}
\begin{align}
	&\begin{aligned}
	\eta_{\msf{D}}^{(3D)}
	=\,
	0.0013
	\ktf^2 \pf \frac{T}{\Tf}
	=\,
	0.0062 \frac{1}{\rs^2 a_0^3} \frac{T}{\Tf},
	\end{aligned}
	\\
	&\begin{aligned}
	\eta_{\msf{D}}^{(2D)}
	=\,&
	0.0022
	\ktf \pf \frac{T}{\Tf}
	=0.0063 \frac{1}{\rs a_0^2} \frac{T}{\Tf}.
	\end{aligned}
\end{align}
\end{subequations}

In Fig.~\ref{fig:R} panel (a) [(b)], we depict the 3D (2D) drag viscosity in units of $a_0^{-3}$ ($a_0^{-2}$) as a function of dimensionless temperature $T/\Tf$.
As shown in this figure, at sufficiently low temperatures $T/\Tf \ll \alpha$ the drag viscosity exhibits a quadratic temperature dependence which becomes linear as the temperature is raised to $T/\Tf \gg \alpha$. Apart from the explicit value of the crossover temperature (which equals $\Tf \alpha$ for drag viscosity and $\vf/d$ for drag resistivity as will become apparent later), this temperature dependence is also noted for the drag resistivity of double-layered 2DEGs~\cite{DragRev,Langevin,Gramila,Solomon,Jauho}.

\subsection{Drag Resistivity}

\begin{figure}[t!]
	\centering
	\includegraphics[width=0.9\linewidth]{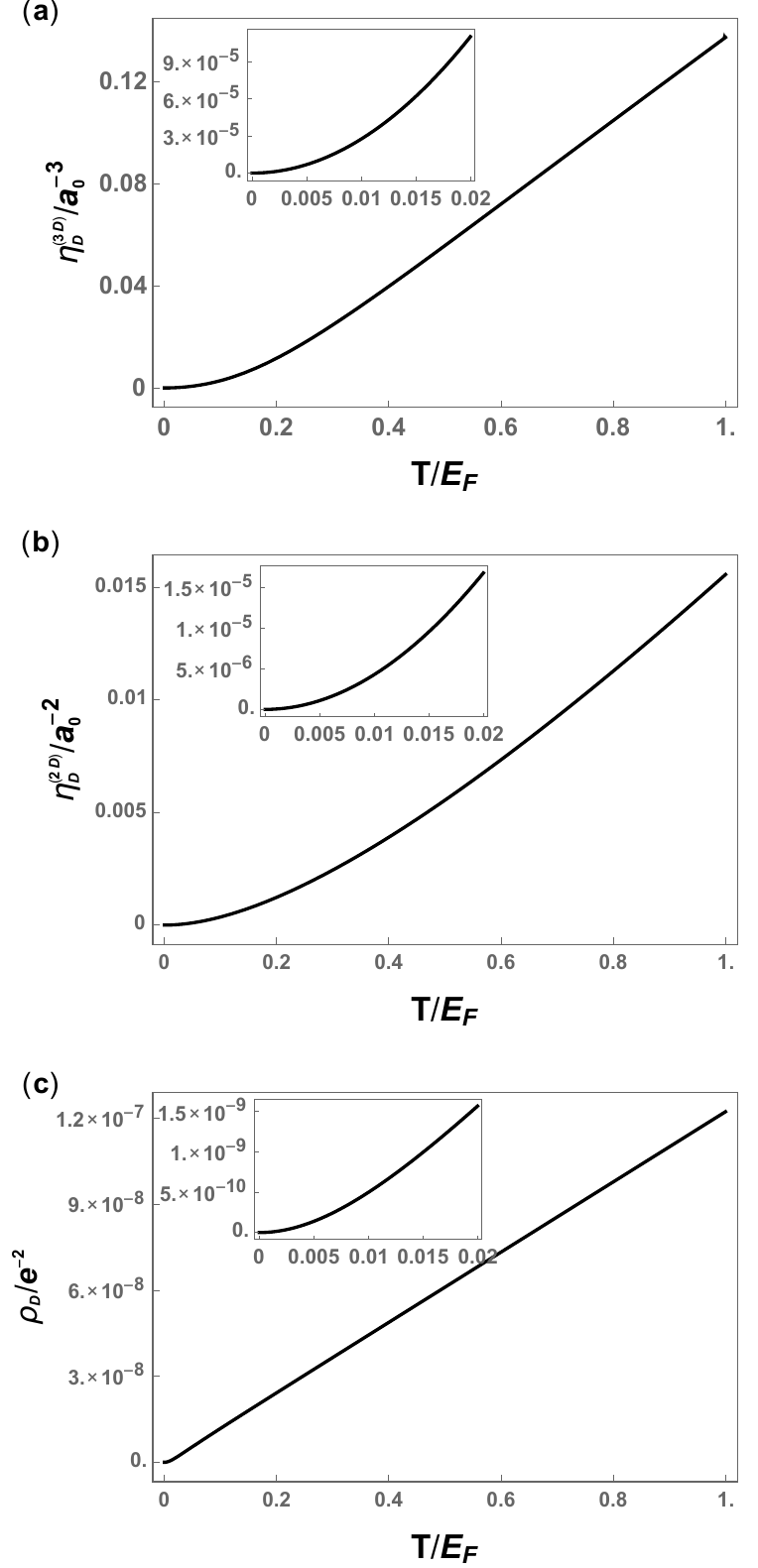}
	\caption{
		Temperature dependence of the drag viscosity $\eta_{\msf{D}}$ and drag resistivity $\rho_{\msf{D}}$.
		Panel (a) shows the drag viscosity of a 3D metal in units of $a_0^{-3}$, panel (b) illustrates the corresponding 2D result in units of $a_0^{-2}$ and panel (c) depicts the drag resistivity of a double-layer 2DEG in units of $e^{-2}$.
		They're obtained from numerical computation of Eqs.~\ref{eq:eta3d}, ~\ref{eq:eta2d} and~\ref{eq:drag-t2} respectively.
		Inset in each panel shows a zoom-in view of the drag viscosity or drag resistivity in the low temperature region.
		Both the drag viscosity and drag resistivity exhibit a crossover from quadratic-in-$T$ behavior at low temperatures and a linear one at higher temperatures.
		In all panels, the dimensionless interaction parameter $\rs$ has been set to be $\rs=0.2$, and in panel (c) the interlayer distance $d$ is set to be $5a_0$.
	}
	\label{fig:R}
\end{figure}

In the following, we will compute the drag resistivity in a manner analogous to that of drag viscosity.
The interlayer distance $d$ is assumed to be much larger than the Thomas-Fermi screening length $\ktf^{-1}$. In this case, the integral in Eq.~\ref{eq:drag-rho} is dominated by momentum $q \lesssim d^{-1}$ and frequency $\ww \lesssim T$. At temperatures $T/\Tf \ll (\pf d)^{-1}$, the main contribution comes from the region $u \ll 1$, and the drag resistivity (Eq.~\ref{eq:drag-R-1}) is well approximated by its static limit ($\ww \ll \vf q$) expression
\begin{align}\label{eq:drag-rho-s}
\begin{aligned}
	\rho_{\msf{D}}
	=
	\, \frac{1}{2n^2 e^2 }
	\int_{\ww,\qb}
	\left[ q_x \tilde{V}_{\msf{inter}}(q) \right]^2
	\dfrac{\left[ \im \Pi^{(R)}(\qb,\ww) \right]^2 }
	{T \sinh^2 \left( \frac{\ww}{2T}\right) },
\end{aligned}
\end{align}
(cf. the analogous static formula Eq.~\ref{eq:drag-eta-s} for drag viscosity).
Here $\tilde{V}(q)$ denotes the static screened interaction and is given by
\begin{align}
	\tilde{V}(q) = \left[ V^{-1}(q) +\nu_0 \IL \right]^{-1}.
\end{align}
For large enough interlayer distance $d \gg \ktf^{-1}$, one has
\begin{align}
\begin{aligned}
	\tilde{V}_{\msf{inter}}(q)
	=\,
	\dfrac{1}{ 2 \nu_0 \ktf } \dfrac{ q}
	{ \sinh {qd} },
\end{aligned}
\end{align}
where $\ktf=2\pi e^2\nu_0$.

Combining Eq.~\ref{eq:drag-rho-s} with the fact that $\im \Pi^{(R)} (\qb,\ww)=-\nu_0 \dfrac{\ww}{\vf q}$ in the static limit $\ww \ll \vf q$, one finds the drag resistivity of a clean double-layer 2DEG at temperatures $T/\Tf \ll 1/\pf d$~\cite{,DragRev,MacDonald,Flensberg,Kamenev}:
\begin{align}\label{eq:drag-t1}
\begin{aligned}
	\rho_{\msf{D}}
	=&
	\frac{1}{2^6\pi^2}
	\dfrac{1}{ n^2 e^2\ktf^2\vf^2 }
	\int_{-\infty}^{\infty}
	d\ww
	\dfrac{\ww^2}
	{T \sinh^2 \left( \frac{\ww}{2T}\right) }
	\int_{\frac{|\ww|}{\vf q}}^{\infty} dq
	\dfrac{ q^3}{ \sinh^2 {qd} } 
	\\
	=&
	\frac{\zeta[3]\pi^2}{16}
	\dfrac{1}{ e^2 \pf^2 \ktf^2 d^4}
	\left( \frac{T}{\Tf}\right)^2
	=
	\frac{\zeta[3]\pi^2}{128}
	\frac{r_s^2}{e^2}
	\left( \frac{a_0}{d}\right)^4
	\left( \frac{T}{\Tf}\right)^2
	.
\end{aligned}
\end{align}

At higher temperatures $T/\Tf  \gtrsim 1/\pf d$, the static limit result (Eq.~\ref{eq:drag-t1}) is no longer valid. Insertion of the expression for the polarization operator $\Pi^{(R)}(u,z)$ at $z=0$ and arbitrary $u$ into Eq.~\ref{eq:drag-R-1}
leads to
\begin{align}\label{eq:drag-t2}
\begin{aligned}
	\rho_{\msf{D}}
	=\,&
	\frac{1}{32 \pi^2}
	\frac{1}{n^2 e^2}
	\frac{\vf}{\ktf^2 d^7} 
	\int_{0}^{\infty} dx 
	\int_{0}^{1} d u
	\dfrac{1}{T \sinh^2 \left( \frac{\vf x u}{2dT}\right)  }
	\\
	\times &
	u^2 \left( 1-u^2 \right) 
	\dfrac{	x^6 }{
	 \sinh^2 x
	}.
\end{aligned}
\end{align}
Here $x\equiv q d$, and we have used the fact that $\ktf d \gg 1$.

Approximating $1/ \sinh^2 \left( \frac{\vf x u}{2dT}\right)$ with $ \left( \frac{\vf x u}{2dT}\right)^{-2}$ in Eq.~\ref{eq:drag-t2}, one finds that the drag resistivity is given by the following expression at temperatures $T/\Tf \gg 1/\pf d$~\cite{Gramila,Solomon,Jauho, Langevin}:
\begin{align}
	\rho_{\msf{D}}
	=\,
	\frac{\pi^4}{180}
	\frac{1}{e^2\ktf^2 \pf^3 d^5} 
	\frac{T}{\Tf}
	=\,
	\frac{\pi^4}{1440\sqrt{2}} \frac{\rs^3}{e^2}
	\left( \frac{a_0}{d}\right)^5 
	\frac{T}{\Tf}.
\end{align}

Using Eq.~\ref{eq:drag-R-1} derived from Kubo formula Eq.~\ref{eq:Kubo-R}, we have proved that the drag resistivity crosses over from the quadratic behavior $\rho_{\msf{D}} \propto T^2$ for temperatures below $\vf/ d$ to a linear one $\rho_{\msf{D}} \propto T$ for temperatures above $\vf/ d$, see panel (c) of Fig.~\ref{fig:R}.

\subsection{Discussion}

As shown in the previous subsections, both the drag viscosity and drag resistivity exhibit a quadratic-in-$T$ behavior at low temperature $T \ll T_c$  and a linear-in-$T$ behavior at higher temperature $T \gg T_c$. Here
the crossover temperature takes the value of $T_c/\Tf=\alpha$ ($\alpha=(16/3\pi^2)^{1/3}\sqrt{\rs}$ in 3D and $\alpha=\sqrt{2} \rs$ in 2D) for the drag viscosity and $T_c/\Tf=1/\pf d$ for the drag resistivity.
It is determined by the boundary of the particle-hole continuum $\ww=\vf q$.

In Eq.~\ref{eq:general}, we show that both the drag viscosity (Eq.~\ref{eq:drag-eta}) and drag resistivity  (Eq.~\ref{eq:drag-R-1}) can be expressed as a two-variable integral with the integrand being a product of the square of the dressed vertex $\tilde{f}(\qb,\ww)$  and phase space factor $R(\qb,\ww)$.
The typical energy tranfer $\ww$ there is of the order of the temperature $T$, while the typical momentum transfer $q$ is of the order of the Thomas-Fermi screening wavevector $\ktf$ for the drag viscosity, and the inverse of interlayer spacing $d^{-1}$ for the drag resistivity (assuming $d \gg \ktf^{-1}$). The crossover temperature $T_c$ is the temperature at which the typical energy  transfer $\ww$
and typical momentum transfer $q$ reach the particle-hole continuum boundary $\ww=\vf q$, and as a result equals $\vf \ktf$ for the drag viscosity and $\vf d^{-1}$ for the drag resistivity.

At temperature $T \ll T_c$, the energy transfer $\ww$ which is of the order of $T$ is much smaller compared with $\vf q$ which is of the order of $T_c$. Therefore, the dressed vertex $\tilde{f}(\qb,\ww)$ in Eq.~\ref{eq:general} can be approximated by $f(\qb)$ with the bare interaction $V(q)$ replaced by the static screened one $\tilde{V}(q)$. Now $\tilde{f}(\qb,\ww)$ becomes frequency independent, and the quadratic temperature dependence of $\eta_{\msf{D}}$ and $\rho_{\msf{D}}$ comes entirely from the phase space $\int_{\ww}R(\qb,\ww)$ of the interlayer scattering with momentum transfer $\qb$.

At temperature $T \gg T_c$, the energy transfer $\ww$ is now cut off by the boundary of the particle-hole continuum $ \vf q$ which is of the order of the crossover temperature $T_c$.
As a result, the factor $\sinh (\ww/2T)$ in the denominator of $R(\qb,\ww)$ (Eq.~\ref{eq:R}) can be approximated by $\ww/2T$ since $\ww \sim T_c \ll T$. Substituting $R(\qb,\ww)$ back into Eq.~\ref{eq:general}, it is now easy to see the linear temperature dependence of the drag viscosity and drag resistivity in this regime.


\section{Conclusion}~\label{sec:conclusion}

In this paper, we show the close resemblance between the drag resistivity of double-layer  2DEGs and the drag viscosity of clean metals in 2D and 3D.
These two phenomena both originate from frictional drag force induced by Coulomb interactions.
Using Kubo formula, we extract the drag resistivity and drag viscosity from the correlation functions of drag force and that of momentum flow induced by drag force, i.e., interacting part of the stress tensor, respectively.
Similar to the drag resistivity, the drag viscosity exhibits a quadratic-in-$T$ scaling behavior at low temperatures which crosses over to a linear one at higher temperatures (which are still low enough such that the plasmon contribution can be ignored). 
It is therefore relatively small compared to the ``Drude''-type viscosity which results from the correlation function of non-interacting stress tensor and has a temperature dependence of $1/T^2$~\cite{abrikosov1957,Abrikosov,pomeranchuk}, 

Even though drag viscosity is relatively small for clean metals, it might be significant for electronic systems with a different Fermi surface geometry, energy dispersion or type of interaction, especially for those with strong Coulomb drag signal.
Another system, where the drag viscosity may have a more pronounced effect is graphene close to charge neutrality in the hydrodynamic regime~\cite{GH-Kumar,GH-Bandurin,GH-Crossno}. It has been found by M\"uller et al.~\cite{Muller} that the ``Drude''-type viscosity has a temperature dependence of $T^2$ at charge neutrality, in contrast to the usual Fermi liquid result of $\eta \propto 1/T^2$~\cite{abrikosov1957,Abrikosov,pomeranchuk}. 
In addition, due to the lack of screening, the interactions become strong at the charge neutrality, resulting in a strong drag force.
As a consequence of particle-hole symmetry, the leading order contribution to the drag resistivity of a double-layer intrinsic graphene vanishes~\cite{GDrag-DasSarma,GDrag-Narozhny}.
However, it has been shown both experimentally and theoretically that application of a magnetic field generates a giant magnetodrag for graphene at charge neutrality~\cite{Mdrag-Levitov,Mdrag-Titov,Mdrag-exp}.
Given the similarity between the drag resistivity and drag viscosity, it is reasonable to expect the drag viscosity might be significant for graphene at charge neutrality in the presence of a  magnetic field. Another type of system where the drag viscosity may dominate are quantum critical systems~\cite{QCP}. The soft fluctuations near the critical point there induce long-range forces, unscreened at criticality, which may give rise to a noticeable viscous drag effect.

In particular, the drag viscosity may also play a role in developing a better understanding of linear-in-$T$ resistivity of the strange metal phase of cuprates~\cite{cuprate}. 
In Ref.~\cite{Davison}, Davison et al. present a mechanism to explain the linear-in-$T$ resistivity by connecting it with the electron viscosity. In particular, they consider a strongly interacting quantum critical system which is weakly coupled to impurities and can be described hydrodynamically, and assume the theory is applicable to actual electronic systems such as cuprates.
For this hydrodynamics liquid, the resistivity is governed by the momentum dissipation characterized by viscosity. 
They found a viscous contribution to resistivity proportional to viscosity $\rho(T) \propto \eta(T)$, see also Refs.~\cite{ElHydro,Spivak} where the same result has been derived. Assuming the lower bound for the ratio of shear viscosity to entropy density~\cite{KSS} is reached, they conclude that the viscous contribution is proportional to the electron entropy $\rho(T) \propto s(T)$ and therefore shows a linear-in-$T$ temperature dependence for strange metals with entropy $s \propto T$~\cite{EntropyCuprate}.
For a critical quantum liquid, the drag viscosity might be dominant and therefore relevant to the linear-in-$T$ resistivity.

In summary, this paper has introduced a novel mechanism for electronic viscosity, based on a drag effect originating from long-range forces. The drag viscosity shows a cross-over from a quadratic $T$-dependence in the $T \to 0$ limit to a linear-in-$T$ behavior, which may result in a linear-in-$T$ conductivity~\cite{Davison,ElHydro,Spivak}. This contribution is shown to be small for regular Fermi liquids with an isotropic Fermi surface. However, we show a nearly one-to-one correspondence between the drag viscosity coefficient and the Coulomb drag resistance, thereby providing guidance in search for systems where the intrinsic drag effect may dominate and give rise to a pronounced linear-in-$T$ behavior of transport coefficients.

\begin{acknowledgements}
	
	The authors are grateful to Dam T. Son, Sergey Syzranov, and Mikhail Titov for useful discussions.  This research was supported by  NSF DMR-1613029 (Y.L.), DOE-BES (DESC0001911) (V.G.), and the Simons Foundation. The authors acknowledge the hospitality of KITP-UCSB, which is supported in part by Grant No. NSF PHY-1748958.
	
\end{acknowledgements}

\vspace{10mm}

\appendix
\appendixheaderon

\section{Diagrammatic Derivation of the Stress Tensor and Drag Force Correlation Functions}\label{app:diagram}

In this Appendix, we provide an alternative diagrammatic approach to calculate the retarded correlation function $\GG_{\tF}^{(R)}$ of stress tensor $\Txy$ and drag force $F^{\LL/\RR}_x$, which are represented by $\tilde{F}$ (Eq.~\ref{eq:tF}) with different forms of vertex $f(\qb)$ (Eq.~\ref{eq:fq}).

The correlation function $\GG_{\tF}^{(R)}$, depicted in Fig.~\ref{fig:E}, is given by the non-interacting correlation function of the bare force $\tF$ and dressed force $\tF^{(d)}$
\begin{align}\label{eq:GF-a1}
	i\GG^{(R)}_{\tF}(\WW)
	=\,
	\braket{\tF_{\msf{cl}}^{(d)}(\WW)\tF_{\msf{q}}(-\WW)}_0
	=\,
	\braket{\tF_{\msf{cl}}(\WW)\tF_{\msf{q}}^{(d)}(-\WW)}_0.
\end{align} 
Here the dressed force $\tilde{F}^{(d)}$ is defined diagrammatically in Fig.~\ref{fig:T}. 
Through a straightfoward calculation, we find the classical components of the dressed $\tilde{F}^{(d)}$ takes the form
\begin{widetext}
\begin{align}\label{eq:dF}
\begin{aligned}
	&\tilde{F}_{\msf{cl}}^{(d)}(\WW)
	=\,
	\frac{1}{2\sqrt{2}}
	\int_{\pb,\pb',\e,\e',\qb,\ww}
	f(\qb)
	V^{-1}_{\LL i'} (q) 
	V^{-1}_{\RR j'} (q) 
	\\
	&
	\left\lbrace 
	\begin{aligned}
	&
	\begin{bmatrix}
	0 & 1
	\end{bmatrix}
	\begin{bmatrix}
	D^{(A)}_{i'i}(\qb,\ww+\WW)
	&
	0
	\\
	D^{(K)}_{i'i}(\qb,\ww+\WW)
	&
	D^{(R)}_{i'i}(\qb,\ww+\WW)
	\end{bmatrix}
	\begin{bmatrix}
	\bar{\psi}^{i}_{\sigma'}  (\pb',\e')	
	\mf(\e') 
	\mf(\e'+\ww+\WW)
	\psi^{i}_{\sigma'}  (\pb'+\qb,\e'+\ww+\WW)	
	\\
	\bar{\psi}^{i}_{\sigma'}  (\pb',\e')	
	\mf(\e') 
	\hat{\tau}^1
	\mf(\e'+\ww+\WW)
	\psi^{i}_{\sigma'}  (\pb'+\qb,\e'+\ww+\WW)
	\end{bmatrix} 
	\\
	\times &
	\vex{u}^{\msf{T}}
	\begin{bmatrix}
	D^{(A)}_{j'j}(-\qb,-\ww)
	&
	0
	\\
	D^{(K)}_{j'j}(-\qb,-\ww)
	&
	D^{(R)}_{j'j}(-\qb,-\ww)
	\end{bmatrix}
	\begin{bmatrix}
	\bar{\psi}^{j}_{\sigma} (\pb+\qb,\e+\ww) 
	\mf(\e+\ww) 
	\mf(\e)
	\psi^{j}_{\sigma}  (\pb,\e)	
	\\
	\bar{\psi}^{j}_{\sigma} (\pb+\qb,\e+\ww) 
	\mf(\e+\ww) 
	\hat{\tau}^1
	\mf(\e)
	\psi^{j}_{\sigma}  (\pb,\e)	
	\end{bmatrix} 
	\\
	+
	&
	\begin{bmatrix}
	1 & 0
	\end{bmatrix}
	\begin{bmatrix}
	D^{(A)}_{i'i}(\qb,\ww+\WW)
	&
	0
	\\
	D^{(K)}_{i'i}(\qb,\ww+\WW)
	&
	D^{(R)}_{i'i}(\qb,\ww+\WW)
	\end{bmatrix}
	\begin{bmatrix}
	\bar{\psi}^{i}_{\sigma'}  (\pb',\e')	
	\mf(\e') 
	\mf(\e'+\ww+\WW)
	\psi^{i}_{\sigma'}  (\pb'+\qb,\e'+\ww+\WW)	
	\\
	\bar{\psi}^{i}_{\sigma'}  (\pb',\e')	
	\mf(\e') 
	\hat{\tau}^1
	\mf(\e'+\ww)
	\psi^{i}_{\sigma'}  (\pb'+\qb,\e'+\ww+\WW)
	\end{bmatrix} 
	\\
	\times	&
	\vex{v}^{\msf{T}}
	\begin{bmatrix}
	D^{(A)}_{j'j}(-\qb,-\ww)
	&
	0
	\\
	D^{(K)}_{j'j}(-\qb,-\ww)
	&
	D^{(R)}_{j'j}(-\qb,-\ww)
	\end{bmatrix}
	\begin{bmatrix}
	\bar{\psi}^{j}_{\sigma} (\pb+\qb,\e+\ww) 
	\mf(\e+\ww) 
	\mf(\e)
	\psi^{j}_{\sigma}  (\pb,\e)	
	\\
	\bar{\psi}^{j}_{\sigma} (\pb+\qb,\e+\ww) 
	\mf(\e+\ww) 
	\hat{\tau}^1
	\mf(\e)
	\psi^{j}_{\sigma}  (\pb,\e)	
	\end{bmatrix}
	\end{aligned}
	\right\rbrace, 
\end{aligned}
\end{align} 
where
$
	\vex{u}^{\msf{T}}
	=\begin{bmatrix}
	0 & 1
	\end{bmatrix}
$
and 
$
	\vex{v}^{\msf{T}}
	=\begin{bmatrix}
	1& 0
	\end{bmatrix}
$.
The quantum component of $\tilde{F}^{(d)}_{\msf{q}}$ can also be obtained from Eq.~\ref{eq:dF} by interchanging $\vex{u}$ and $\vex{v}$.

Insertion of Eqs.~\ref{eq:dF} and~\ref{eq:tF2} into Eq.~\ref{eq:GF-a1} yields the retarded correlation function of $\tF$,
\begin{align}\label{eq:GF3}
\begin{aligned}
	&i\GG^{(R)}_{\tF}(\WW)
	=\,
	-\frac{1}{2}
	\int_{\ww,\qb}
	\suml{i=\LL,\RR}
	 f(\qb) f(-\zeta_i \qb)
	 \left\lbrace 
	\left[
	\vex{u}^{\msf{T}} W_{\LL i}(\qb,\ww+\WW)\vex{u}
	 \right] 
	  \left[
	  \vex{u}^{\msf{T}} W_{\RR \bar{i}}(-\qb,-\ww)\vex{v}
	  \right] 
	  +
	  \left[
	  \vex{v}^{\msf{T}} W_{\LL i}(\qb,\ww+\WW)\vex{v}
	  \right] 
	  \left[
	  \vex{v}^{\msf{T}} W_{\RR \bar{i}}(-\qb,-\ww)\vex{u}
	  \right] 
	  \right. 
	  \\
	  & \left. 
	  +
	  \left[
	  \vex{u}^{\msf{T}} W_{\LL i}(\qb,\ww+\WW)\vex{v}
	  \right] 
	  \left[
	  \vex{u}^{\msf{T}} W_{\RR \bar{i}}(-\qb,-\ww)\vex{u}
	  \right] 
	  +
	  \left[
	  \vex{v}^{\msf{T}} W_{\LL i}(\qb,\ww+\WW)\vex{u}
	  \right] 
	  \left[
	  \vex{v}^{\msf{T}} W_{\RR \bar{i}}(-\qb,-\ww)\vex{v}
	  \right] 
	  \right\rbrace,
\end{aligned}
\end{align}
\end{widetext}
where $W$ is defined as
\begin{align}
\begin{aligned}
	&W(\qb,\ww)
	\equiv
	 V^{-1}(q)
	\htau^1_{\msf{L}}
	D({\qb,\ww})
	\Pi({\qb,\ww}).
\end{aligned}
\end{align}
$\htau^1_{\msf{L}}$ is the Pauli matrix acting on the layer space.

Using the causality structures of dressed propagator $D(\qb,\ww)$ (Eq.~\ref{eq:D2}) and self energy $\Pi (\qb,\ww)$ (Eq.~\ref{eq:Pi1}) of the bosonic field $\phi$, we find Eq.~\ref{eq:GF3} can be further simplified to Eq.~\ref{eq:GF-2} which is derived in Sec.~\ref{sec:Cor} through a different approach that involves a source field.

\section{Zero Temperature Polarization Operator}\label{app:Pi}

The non-interacting polarization operator $\Pi^{(R)} (\qb,\ww)$ at zero temperature was calculated by Lindhard~\cite{lindhard} and Stern~\cite{stern} respectively for a Fermi liquid in 2D and 3D. In this appendix, we give the explicit expressions which are needed for our calculations of drag viscosity and drag resistivity.

In terms of dimensionless parameter $u \equiv \ww/\vf q$ and $z=q/2\pf$, the real and imaginary parts of polarization operator $\Pi(u,z)$ for $u>0$ in 3D are given by, respectively,
	\begin{align}
	&\begin{aligned}\label{eq:re-pi-3}
		\re \Pi^{(R)} (u,z)
		=
		-\nu_0
		&
		\left\lbrace 
		\frac{1}{2}
		+\frac{1}{8z} \left[ 1-\left( z-u \right)^2\right]  \ln \bigg\lvert \frac{z-u+1}{z-u-1} \bigg\rvert 
		\right. 
		\\
		&\, \, \left. +\frac{1}{8z} \left[ 1-\left( z+u \right)^2\right]  \ln \bigg\lvert \frac{z+u+1}{z+u-1} \bigg\rvert
		\right\rbrace,  
	\end{aligned}	
	\end{align}
	\begin{align}
	&\begin{aligned}\label{eq:im-pi-3}
	&\im \Pi^{(R)} (u,z)
	=
	-\pi \nu_0
	\left\lbrace 
	\frac{1}{2} u \Theta \left( \lvert 1-z\rvert - u \right) 
	\right. 
	\\
	& \,\, \left. +
	\frac{1}{8z} \left[ 1-\left( z-u \right)^2\right] \Theta ( 1+z-u) \Theta\left( u-\lvert 1-z \rvert\right) 
	\right\rbrace.
	\end{aligned}	
	\end{align}
In 2D, the polarization operator $\Pi^{(R)}(u>0,z)$ acquires the form
	\begin{align}
	&\begin{aligned}\label{eq:re-pi-2}
	&\re \Pi^{(R)} (u,z)
	=
	-\nu_0
	\left\lbrace 
	1
	\right. 
	\\
	&-
	\frac{\sgn{(z+u)}}{2z}
	\sqrt{\left( z+u\right)^2-1}
	\,
	\Theta \left[ \lvert z+u \rvert-1\right] 
	 \\
	&\,\,\left. -
	\frac{\sgn{(z-u)}}{2z}
	\sqrt{\left( z-u\right)^2-1}
	\,
	\Theta \left[ \lvert z-u \rvert-1\right] 
	\right\rbrace ,
	\end{aligned}
	\end{align}
	\begin{align}
	\begin{aligned}\label{eq:im-pi-2}
	&\im \Pi^{(R)} (u,z)
	=
	\nu_0 \frac{1}{2z}
	\left\lbrace 
	\sqrt{1-\left( z+u\right)^2}
	\Theta \left[1- \lvert z+u \rvert\right] 
	\right. 
	\\
	&\,\,\left. 
	-\sqrt{1-\left( z-u\right)^2}
	\Theta \left[ 1-\lvert z-u \rvert\right] 
	\right\rbrace .
	\end{aligned}
	\end{align}
For $u<0$, the explicit expressions for the polarization operator can be deduced from the equations above using $\im \Pi (u,x)=- \im \Pi(-u,x)$ and  $\re \Pi (u,x)=\re \Pi (-u,x)$.

\bibliography{Drag.bib}

\begin{thebibliography}{72}
\expandafter\ifx\csname natexlab\endcsname\relax\def\natexlab#1{#1}\fi
\expandafter\ifx\csname bibnamefont\endcsname\relax
  \def\bibnamefont#1{#1}\fi
\expandafter\ifx\csname bibfnamefont\endcsname\relax
  \def\bibfnamefont#1{#1}\fi
\expandafter\ifx\csname citenamefont\endcsname\relax
  \def\citenamefont#1{#1}\fi
\expandafter\ifx\csname url\endcsname\relax
  \def\url#1{\texttt{#1}}\fi
\expandafter\ifx\csname urlprefix\endcsname\relax\def\urlprefix{URL }\fi
\providecommand{\bibinfo}[2]{#2}
\providecommand{\eprint}[2][]{\url{#2}}

\bibitem[{\citenamefont{Andreev et~al.}(2011)\citenamefont{Andreev, Kivelson,
  and Spivak}}]{ElHydro}
\bibinfo{author}{\bibfnamefont{A.~V.} \bibnamefont{Andreev}},
  \bibinfo{author}{\bibfnamefont{S.~A.} \bibnamefont{Kivelson}},
  \bibnamefont{and} \bibinfo{author}{\bibfnamefont{B.}~\bibnamefont{Spivak}},
  \bibinfo{journal}{Phys. Rev. Lett.} \textbf{\bibinfo{volume}{106}},
  \bibinfo{pages}{256804} (\bibinfo{year}{2011}).

\bibitem[{\citenamefont{Narozhny et~al.}(2017)\citenamefont{Narozhny, Gornyi,
  Mirlin, and Schmalian}}]{GHydro-1}
\bibinfo{author}{\bibfnamefont{B.~N.} \bibnamefont{Narozhny}},
  \bibinfo{author}{\bibfnamefont{I.~V.} \bibnamefont{Gornyi}},
  \bibinfo{author}{\bibfnamefont{A.~D.} \bibnamefont{Mirlin}},
  \bibnamefont{and}
  \bibinfo{author}{\bibfnamefont{J.}~\bibnamefont{Schmalian}},
  \bibinfo{journal}{Annalen der Physik} \textbf{\bibinfo{volume}{529}},
  \bibinfo{pages}{1700043} (\bibinfo{year}{2017}).

\bibitem[{\citenamefont{Lucas et~al.}(2016)\citenamefont{Lucas, Crossno, Fong,
  Kim, and Sachdev}}]{GHydro-2}
\bibinfo{author}{\bibfnamefont{A.}~\bibnamefont{Lucas}},
  \bibinfo{author}{\bibfnamefont{J.}~\bibnamefont{Crossno}},
  \bibinfo{author}{\bibfnamefont{K.~C.} \bibnamefont{Fong}},
  \bibinfo{author}{\bibfnamefont{P.}~\bibnamefont{Kim}}, \bibnamefont{and}
  \bibinfo{author}{\bibfnamefont{S.}~\bibnamefont{Sachdev}},
  \bibinfo{journal}{Phys. Rev. B} \textbf{\bibinfo{volume}{93}},
  \bibinfo{pages}{075426} (\bibinfo{year}{2016}).

\bibitem[{\citenamefont{Lucas and Fong}(2018)}]{GHydro-3}
\bibinfo{author}{\bibfnamefont{A.}~\bibnamefont{Lucas}} \bibnamefont{and}
  \bibinfo{author}{\bibfnamefont{K.~C.} \bibnamefont{Fong}},
  \bibinfo{journal}{Journal of Physics: Condensed Matter}
  \textbf{\bibinfo{volume}{30}}, \bibinfo{pages}{053001}
  (\bibinfo{year}{2018}).

\bibitem[{\citenamefont{M\"uller et~al.}(2009)\citenamefont{M\"uller,
  Schmalian, and Fritz}}]{Muller}
\bibinfo{author}{\bibfnamefont{M.}~\bibnamefont{M\"uller}},
  \bibinfo{author}{\bibfnamefont{J.}~\bibnamefont{Schmalian}},
  \bibnamefont{and} \bibinfo{author}{\bibfnamefont{L.}~\bibnamefont{Fritz}},
  \bibinfo{journal}{Phys. Rev. Lett.} \textbf{\bibinfo{volume}{103}},
  \bibinfo{pages}{025301} (\bibinfo{year}{2009}).

\bibitem[{\citenamefont{Torre et~al.}(2015)\citenamefont{Torre, Tomadin, Geim,
  and Polini}}]{GrapheneV-1}
\bibinfo{author}{\bibfnamefont{I.}~\bibnamefont{Torre}},
  \bibinfo{author}{\bibfnamefont{A.}~\bibnamefont{Tomadin}},
  \bibinfo{author}{\bibfnamefont{A.~K.} \bibnamefont{Geim}}, \bibnamefont{and}
  \bibinfo{author}{\bibfnamefont{M.}~\bibnamefont{Polini}},
  \bibinfo{journal}{Phys. Rev. B} \textbf{\bibinfo{volume}{92}},
  \bibinfo{pages}{165433} (\bibinfo{year}{2015}).

\bibitem[{\citenamefont{Levitov and Falkovich}(2016)}]{GrapheneV-2}
\bibinfo{author}{\bibfnamefont{L.}~\bibnamefont{Levitov}} \bibnamefont{and}
  \bibinfo{author}{\bibfnamefont{G.}~\bibnamefont{Falkovich}},
  \bibinfo{journal}{Nature Physics} \textbf{\bibinfo{volume}{12}},
  \bibinfo{pages}{672} (\bibinfo{year}{2016}).

\bibitem[{\citenamefont{Principi et~al.}(2016)\citenamefont{Principi, Vignale,
  Carrega, and Polini}}]{GrapheneV-3}
\bibinfo{author}{\bibfnamefont{A.}~\bibnamefont{Principi}},
  \bibinfo{author}{\bibfnamefont{G.}~\bibnamefont{Vignale}},
  \bibinfo{author}{\bibfnamefont{M.}~\bibnamefont{Carrega}}, \bibnamefont{and}
  \bibinfo{author}{\bibfnamefont{M.}~\bibnamefont{Polini}},
  \bibinfo{journal}{Physical Review B} \textbf{\bibinfo{volume}{93}},
  \bibinfo{pages}{125410} (\bibinfo{year}{2016}).

\bibitem[{\citenamefont{Briskot et~al.}(2015)\citenamefont{Briskot, Sch{\"u}tt,
  Gornyi, Titov, Narozhny, and Mirlin}}]{GrapheneV-4}
\bibinfo{author}{\bibfnamefont{U.}~\bibnamefont{Briskot}},
  \bibinfo{author}{\bibfnamefont{M.}~\bibnamefont{Sch{\"u}tt}},
  \bibinfo{author}{\bibfnamefont{I.~V.} \bibnamefont{Gornyi}},
  \bibinfo{author}{\bibfnamefont{M.}~\bibnamefont{Titov}},
  \bibinfo{author}{\bibfnamefont{B.~N.} \bibnamefont{Narozhny}},
  \bibnamefont{and} \bibinfo{author}{\bibfnamefont{A.~D.}
  \bibnamefont{Mirlin}}, \bibinfo{journal}{Physical Review B}
  \textbf{\bibinfo{volume}{92}}, \bibinfo{pages}{115426}
  (\bibinfo{year}{2015}).

\bibitem[{\citenamefont{Sherafati et~al.}(2016)\citenamefont{Sherafati,
  Principi, and Vignale}}]{Hall}
\bibinfo{author}{\bibfnamefont{M.}~\bibnamefont{Sherafati}},
  \bibinfo{author}{\bibfnamefont{A.}~\bibnamefont{Principi}}, \bibnamefont{and}
  \bibinfo{author}{\bibfnamefont{G.}~\bibnamefont{Vignale}},
  \bibinfo{journal}{Phys. Rev. B} \textbf{\bibinfo{volume}{94}},
  \bibinfo{pages}{125427} (\bibinfo{year}{2016}).

\bibitem[{\citenamefont{Scaffidi et~al.}(2017)\citenamefont{Scaffidi, Nandi,
  Schmidt, Mackenzie, and Moore}}]{Hall-2}
\bibinfo{author}{\bibfnamefont{T.}~\bibnamefont{Scaffidi}},
  \bibinfo{author}{\bibfnamefont{N.}~\bibnamefont{Nandi}},
  \bibinfo{author}{\bibfnamefont{B.}~\bibnamefont{Schmidt}},
  \bibinfo{author}{\bibfnamefont{A.~P.} \bibnamefont{Mackenzie}},
  \bibnamefont{and} \bibinfo{author}{\bibfnamefont{J.~E.} \bibnamefont{Moore}},
  \bibinfo{journal}{Phys. Rev. Lett.} \textbf{\bibinfo{volume}{118}},
  \bibinfo{pages}{226601} (\bibinfo{year}{2017}).

\bibitem[{\citenamefont{Pellegrino et~al.}(2016)\citenamefont{Pellegrino,
  Torre, Geim, and Polini}}]{EH-1}
\bibinfo{author}{\bibfnamefont{F.~M.~D.} \bibnamefont{Pellegrino}},
  \bibinfo{author}{\bibfnamefont{I.}~\bibnamefont{Torre}},
  \bibinfo{author}{\bibfnamefont{A.~K.} \bibnamefont{Geim}}, \bibnamefont{and}
  \bibinfo{author}{\bibfnamefont{M.}~\bibnamefont{Polini}},
  \bibinfo{journal}{Phys. Rev. B} \textbf{\bibinfo{volume}{94}},
  \bibinfo{pages}{155414} (\bibinfo{year}{2016}).

\bibitem[{\citenamefont{Guo et~al.}(2017)\citenamefont{Guo, Ilseven, Falkovich,
  and Levitov}}]{EH-2}
\bibinfo{author}{\bibfnamefont{H.}~\bibnamefont{Guo}},
  \bibinfo{author}{\bibfnamefont{E.}~\bibnamefont{Ilseven}},
  \bibinfo{author}{\bibfnamefont{G.}~\bibnamefont{Falkovich}},
  \bibnamefont{and} \bibinfo{author}{\bibfnamefont{L.~S.}
  \bibnamefont{Levitov}}, \bibinfo{journal}{Proceedings of the National Academy
  of Sciences} \textbf{\bibinfo{volume}{114}}, \bibinfo{pages}{3068}
  (\bibinfo{year}{2017}).

\bibitem[{\citenamefont{Levchenko et~al.}(2017)\citenamefont{Levchenko, Xie,
  and Andreev}}]{EH-3}
\bibinfo{author}{\bibfnamefont{A.}~\bibnamefont{Levchenko}},
  \bibinfo{author}{\bibfnamefont{H.-Y.} \bibnamefont{Xie}}, \bibnamefont{and}
  \bibinfo{author}{\bibfnamefont{A.~V.} \bibnamefont{Andreev}},
  \bibinfo{journal}{Phys. Rev. B} \textbf{\bibinfo{volume}{95}},
  \bibinfo{pages}{121301(R)} (\bibinfo{year}{2017}).

\bibitem[{\citenamefont{Alekseev}(2016)}]{EH-4}
\bibinfo{author}{\bibfnamefont{P.~S.} \bibnamefont{Alekseev}},
  \bibinfo{journal}{Phys. Rev. Lett.} \textbf{\bibinfo{volume}{117}},
  \bibinfo{pages}{166601} (\bibinfo{year}{2016}).

\bibitem[{\citenamefont{Lucas and Das~Sarma}(2018)}]{EH-5}
\bibinfo{author}{\bibfnamefont{A.}~\bibnamefont{Lucas}} \bibnamefont{and}
  \bibinfo{author}{\bibfnamefont{S.}~\bibnamefont{Das~Sarma}},
  \bibinfo{journal}{Physical Review B} \textbf{\bibinfo{volume}{97}},
  \bibinfo{pages}{245128} (\bibinfo{year}{2018}).

\bibitem[{\citenamefont{Cook and Lucas}(2019)}]{EH-6}
\bibinfo{author}{\bibfnamefont{C.~Q.} \bibnamefont{Cook}} \bibnamefont{and}
  \bibinfo{author}{\bibfnamefont{A.}~\bibnamefont{Lucas}},
  \bibinfo{journal}{Phys. Rev. B} \textbf{\bibinfo{volume}{99}},
  \bibinfo{pages}{235148} (\bibinfo{year}{2019}).

\bibitem[{\citenamefont{Apostolov et~al.}(2014)\citenamefont{Apostolov,
  Levchenko, and Andreev}}]{Apostolov}
\bibinfo{author}{\bibfnamefont{S.~S.} \bibnamefont{Apostolov}},
  \bibinfo{author}{\bibfnamefont{A.}~\bibnamefont{Levchenko}},
  \bibnamefont{and} \bibinfo{author}{\bibfnamefont{A.~V.}
  \bibnamefont{Andreev}}, \bibinfo{journal}{Phys. Rev. B}
  \textbf{\bibinfo{volume}{89}}, \bibinfo{pages}{121104}
  (\bibinfo{year}{2014}).

\bibitem[{\citenamefont{Apostolov et~al.}(2019)\citenamefont{Apostolov, Pesin,
  and Levchenko}}]{Apostolov2}
\bibinfo{author}{\bibfnamefont{S.~S.} \bibnamefont{Apostolov}},
  \bibinfo{author}{\bibfnamefont{D.~A.} \bibnamefont{Pesin}}, \bibnamefont{and}
  \bibinfo{author}{\bibfnamefont{A.}~\bibnamefont{Levchenko}},
  \bibinfo{journal}{Phys. Rev. B} \textbf{\bibinfo{volume}{100}},
  \bibinfo{pages}{115401} (\bibinfo{year}{2019}).

\bibitem[{\citenamefont{Liao and Galitski}(2019)}]{VS}
\bibinfo{author}{\bibfnamefont{Y.}~\bibnamefont{Liao}} \bibnamefont{and}
  \bibinfo{author}{\bibfnamefont{V.}~\bibnamefont{Galitski}},
  \bibinfo{journal}{Physical Review B} \textbf{\bibinfo{volume}{100}},
  \bibinfo{pages}{060501(R)} (\bibinfo{year}{2019}).

\bibitem[{\citenamefont{Kumar et~al.}(2017)\citenamefont{Kumar, Bandurin,
  Pellegrino, Cao, Principi, Guo, Auton, Shalom, Ponomarenko, Falkovich
  et~al.}}]{GH-Kumar}
\bibinfo{author}{\bibfnamefont{R.~K.} \bibnamefont{Kumar}},
  \bibinfo{author}{\bibfnamefont{D.~A.} \bibnamefont{Bandurin}},
  \bibinfo{author}{\bibfnamefont{F.~M.~D.} \bibnamefont{Pellegrino}},
  \bibinfo{author}{\bibfnamefont{Y.}~\bibnamefont{Cao}},
  \bibinfo{author}{\bibfnamefont{A.}~\bibnamefont{Principi}},
  \bibinfo{author}{\bibfnamefont{H.}~\bibnamefont{Guo}},
  \bibinfo{author}{\bibfnamefont{G.}~\bibnamefont{Auton}},
  \bibinfo{author}{\bibfnamefont{M.~B.} \bibnamefont{Shalom}},
  \bibinfo{author}{\bibfnamefont{L.~A.} \bibnamefont{Ponomarenko}},
  \bibinfo{author}{\bibfnamefont{G.}~\bibnamefont{Falkovich}},
  \bibnamefont{et~al.}, \bibinfo{journal}{Nature Physics}
  \textbf{\bibinfo{volume}{13}}, \bibinfo{pages}{1182} (\bibinfo{year}{2017}).

\bibitem[{\citenamefont{Bandurin et~al.}(2016)\citenamefont{Bandurin, Torre,
  Kumar, Shalom, Tomadin, Principi, Auton, Khestanova, Novoselov, Grigorieva
  et~al.}}]{GH-Bandurin}
\bibinfo{author}{\bibfnamefont{D.~A.} \bibnamefont{Bandurin}},
  \bibinfo{author}{\bibfnamefont{I.}~\bibnamefont{Torre}},
  \bibinfo{author}{\bibfnamefont{R.~K.} \bibnamefont{Kumar}},
  \bibinfo{author}{\bibfnamefont{M.~B.} \bibnamefont{Shalom}},
  \bibinfo{author}{\bibfnamefont{A.}~\bibnamefont{Tomadin}},
  \bibinfo{author}{\bibfnamefont{A.}~\bibnamefont{Principi}},
  \bibinfo{author}{\bibfnamefont{G.}~\bibnamefont{Auton}},
  \bibinfo{author}{\bibfnamefont{E.}~\bibnamefont{Khestanova}},
  \bibinfo{author}{\bibfnamefont{K.}~\bibnamefont{Novoselov}},
  \bibinfo{author}{\bibfnamefont{I.}~\bibnamefont{Grigorieva}},
  \bibnamefont{et~al.}, \bibinfo{journal}{Science}
  \textbf{\bibinfo{volume}{351}}, \bibinfo{pages}{1055} (\bibinfo{year}{2016}).

\bibitem[{\citenamefont{Crossno et~al.}(2016)\citenamefont{Crossno, Shi, Wang,
  Liu, Harzheim, Lucas, Sachdev, Kim, Taniguchi, Watanabe et~al.}}]{GH-Crossno}
\bibinfo{author}{\bibfnamefont{J.}~\bibnamefont{Crossno}},
  \bibinfo{author}{\bibfnamefont{J.~K.} \bibnamefont{Shi}},
  \bibinfo{author}{\bibfnamefont{K.}~\bibnamefont{Wang}},
  \bibinfo{author}{\bibfnamefont{X.}~\bibnamefont{Liu}},
  \bibinfo{author}{\bibfnamefont{A.}~\bibnamefont{Harzheim}},
  \bibinfo{author}{\bibfnamefont{A.}~\bibnamefont{Lucas}},
  \bibinfo{author}{\bibfnamefont{S.}~\bibnamefont{Sachdev}},
  \bibinfo{author}{\bibfnamefont{P.}~\bibnamefont{Kim}},
  \bibinfo{author}{\bibfnamefont{T.}~\bibnamefont{Taniguchi}},
  \bibinfo{author}{\bibfnamefont{K.}~\bibnamefont{Watanabe}},
  \bibnamefont{et~al.}, \bibinfo{journal}{Science}
  \textbf{\bibinfo{volume}{351}}, \bibinfo{pages}{1058} (\bibinfo{year}{2016}).

\bibitem[{\citenamefont{Gooth et~al.}(2017{\natexlab{a}})\citenamefont{Gooth,
  Niemann, Meng, Grushin, Landsteiner, Gotsmann, Menges, Schmidt, Shekhar,
  S{\"u}{\ss} et~al.}}]{Weyl}
\bibinfo{author}{\bibfnamefont{J.}~\bibnamefont{Gooth}},
  \bibinfo{author}{\bibfnamefont{A.~C.} \bibnamefont{Niemann}},
  \bibinfo{author}{\bibfnamefont{T.}~\bibnamefont{Meng}},
  \bibinfo{author}{\bibfnamefont{A.~G.} \bibnamefont{Grushin}},
  \bibinfo{author}{\bibfnamefont{K.}~\bibnamefont{Landsteiner}},
  \bibinfo{author}{\bibfnamefont{B.}~\bibnamefont{Gotsmann}},
  \bibinfo{author}{\bibfnamefont{F.}~\bibnamefont{Menges}},
  \bibinfo{author}{\bibfnamefont{M.}~\bibnamefont{Schmidt}},
  \bibinfo{author}{\bibfnamefont{C.}~\bibnamefont{Shekhar}},
  \bibinfo{author}{\bibfnamefont{V.}~\bibnamefont{S{\"u}{\ss}}},
  \bibnamefont{et~al.}, \bibinfo{journal}{Nature}
  \textbf{\bibinfo{volume}{547}}, \bibinfo{pages}{324}
  (\bibinfo{year}{2017}{\natexlab{a}}).

\bibitem[{\citenamefont{Gooth et~al.}(2017{\natexlab{b}})\citenamefont{Gooth,
  Menges, Shekhar, S{\"u}ss, Kumar, Sun, Drechsler, Zierold, Felser, and
  Gotsmann}}]{WP2}
\bibinfo{author}{\bibfnamefont{J.}~\bibnamefont{Gooth}},
  \bibinfo{author}{\bibfnamefont{F.}~\bibnamefont{Menges}},
  \bibinfo{author}{\bibfnamefont{C.}~\bibnamefont{Shekhar}},
  \bibinfo{author}{\bibfnamefont{V.}~\bibnamefont{S{\"u}ss}},
  \bibinfo{author}{\bibfnamefont{N.}~\bibnamefont{Kumar}},
  \bibinfo{author}{\bibfnamefont{Y.}~\bibnamefont{Sun}},
  \bibinfo{author}{\bibfnamefont{U.}~\bibnamefont{Drechsler}},
  \bibinfo{author}{\bibfnamefont{R.}~\bibnamefont{Zierold}},
  \bibinfo{author}{\bibfnamefont{C.}~\bibnamefont{Felser}}, \bibnamefont{and}
  \bibinfo{author}{\bibfnamefont{B.}~\bibnamefont{Gotsmann}},
  \bibinfo{journal}{arXiv preprint arXiv:1706.05925}
  (\bibinfo{year}{2017}{\natexlab{b}}).

\bibitem[{\citenamefont{Moll et~al.}(2016)\citenamefont{Moll, Kushwaha, Nandi,
  Schmidt, and Mackenzie}}]{PdCoO2}
\bibinfo{author}{\bibfnamefont{P.~J.} \bibnamefont{Moll}},
  \bibinfo{author}{\bibfnamefont{P.}~\bibnamefont{Kushwaha}},
  \bibinfo{author}{\bibfnamefont{N.}~\bibnamefont{Nandi}},
  \bibinfo{author}{\bibfnamefont{B.}~\bibnamefont{Schmidt}}, \bibnamefont{and}
  \bibinfo{author}{\bibfnamefont{A.~P.} \bibnamefont{Mackenzie}},
  \bibinfo{journal}{Science} \textbf{\bibinfo{volume}{351}},
  \bibinfo{pages}{1061} (\bibinfo{year}{2016}).

\bibitem[{\citenamefont{Fu et~al.}(2018)\citenamefont{Fu, Scaffidi, Waissman,
  Sun, Saha, Watzman, Srivastava, Li, Schnelle, Werner et~al.}}]{PtSn4}
\bibinfo{author}{\bibfnamefont{C.}~\bibnamefont{Fu}},
  \bibinfo{author}{\bibfnamefont{T.}~\bibnamefont{Scaffidi}},
  \bibinfo{author}{\bibfnamefont{J.}~\bibnamefont{Waissman}},
  \bibinfo{author}{\bibfnamefont{Y.}~\bibnamefont{Sun}},
  \bibinfo{author}{\bibfnamefont{R.}~\bibnamefont{Saha}},
  \bibinfo{author}{\bibfnamefont{S.~J.} \bibnamefont{Watzman}},
  \bibinfo{author}{\bibfnamefont{A.~K.} \bibnamefont{Srivastava}},
  \bibinfo{author}{\bibfnamefont{G.}~\bibnamefont{Li}},
  \bibinfo{author}{\bibfnamefont{W.}~\bibnamefont{Schnelle}},
  \bibinfo{author}{\bibfnamefont{P.}~\bibnamefont{Werner}},
  \bibnamefont{et~al.}, \bibinfo{journal}{arXiv preprint arXiv:1802.09468}
  (\bibinfo{year}{2018}).

\bibitem[{\citenamefont{Gusev et~al.}(2018)\citenamefont{Gusev, Levin,
  Levinson, and Bakarov}}]{GaAs}
\bibinfo{author}{\bibfnamefont{G.}~\bibnamefont{Gusev}},
  \bibinfo{author}{\bibfnamefont{A.}~\bibnamefont{Levin}},
  \bibinfo{author}{\bibfnamefont{E.}~\bibnamefont{Levinson}}, \bibnamefont{and}
  \bibinfo{author}{\bibfnamefont{A.}~\bibnamefont{Bakarov}},
  \bibinfo{journal}{AIP Advances} \textbf{\bibinfo{volume}{8}},
  \bibinfo{pages}{025318} (\bibinfo{year}{2018}).

\bibitem[{\citenamefont{Sch{\"a}fer and Teaney}(2009)}]{VisRev}
\bibinfo{author}{\bibfnamefont{T.}~\bibnamefont{Sch{\"a}fer}} \bibnamefont{and}
  \bibinfo{author}{\bibfnamefont{D.}~\bibnamefont{Teaney}},
  \bibinfo{journal}{Reports on Progress in Physics}
  \textbf{\bibinfo{volume}{72}}, \bibinfo{pages}{126001}
  (\bibinfo{year}{2009}).

\bibitem[{\citenamefont{Kovtun et~al.}(2005)\citenamefont{Kovtun, Son, and
  Starinets}}]{KSS}
\bibinfo{author}{\bibfnamefont{P.~K.} \bibnamefont{Kovtun}},
  \bibinfo{author}{\bibfnamefont{D.~T.} \bibnamefont{Son}}, \bibnamefont{and}
  \bibinfo{author}{\bibfnamefont{A.~O.} \bibnamefont{Starinets}},
  \bibinfo{journal}{Phys. Rev. Lett.} \textbf{\bibinfo{volume}{94}},
  \bibinfo{pages}{111601} (\bibinfo{year}{2005}).

\bibitem[{\citenamefont{Pomeranchuk}(1950)}]{pomeranchuk}
\bibinfo{author}{\bibfnamefont{I.}~\bibnamefont{Pomeranchuk}},
  \bibinfo{journal}{J. Exptl. Theoret. Phys. USSR}
  \textbf{\bibinfo{volume}{20}}, \bibinfo{pages}{919} (\bibinfo{year}{1950}).

\bibitem[{\citenamefont{Abrikosov and Khalatnikov}(1959)}]{Abrikosov}
\bibinfo{author}{\bibfnamefont{A.~A.} \bibnamefont{Abrikosov}}
  \bibnamefont{and} \bibinfo{author}{\bibfnamefont{I.~M.}
  \bibnamefont{Khalatnikov}}, \bibinfo{journal}{Reports on Progress in Physics}
  \textbf{\bibinfo{volume}{22}}, \bibinfo{pages}{329} (\bibinfo{year}{1959}).

\bibitem[{\citenamefont{Abrikosov and Khalatnikov}(1957)}]{abrikosov1957}
\bibinfo{author}{\bibfnamefont{A.~A.} \bibnamefont{Abrikosov}}
  \bibnamefont{and} \bibinfo{author}{\bibfnamefont{I.~M.}
  \bibnamefont{Khalatnikov}}, \bibinfo{journal}{J. Exptl. Theoret. Phys. USSR}
  \textbf{\bibinfo{volume}{32}}, \bibinfo{pages}{1083} (\bibinfo{year}{1957}).

\bibitem[{\citenamefont{Zinov'eva}(1958)}]{He3}
\bibinfo{author}{\bibfnamefont{K.}~\bibnamefont{Zinov'eva}},
  \bibinfo{journal}{J. Exptl. Theoret. Phys.(USSR)}
  \textbf{\bibinfo{volume}{34}}, \bibinfo{pages}{609} (\bibinfo{year}{1958}).

\bibitem[{\citenamefont{Nishida and Son}(2007)}]{Nishida}
\bibinfo{author}{\bibfnamefont{Y.}~\bibnamefont{Nishida}} \bibnamefont{and}
  \bibinfo{author}{\bibfnamefont{D.~T.} \bibnamefont{Son}},
  \bibinfo{journal}{Phys. Rev. D} \textbf{\bibinfo{volume}{76}},
  \bibinfo{pages}{086004} (\bibinfo{year}{2007}).

\bibitem[{\citenamefont{Link et~al.}(2018)\citenamefont{Link, Sheehy, Narozhny,
  and Schmalian}}]{Schmalian}
\bibinfo{author}{\bibfnamefont{J.~M.} \bibnamefont{Link}},
  \bibinfo{author}{\bibfnamefont{D.~E.} \bibnamefont{Sheehy}},
  \bibinfo{author}{\bibfnamefont{B.~N.} \bibnamefont{Narozhny}},
  \bibnamefont{and}
  \bibinfo{author}{\bibfnamefont{J.}~\bibnamefont{Schmalian}},
  \bibinfo{journal}{Physical Review B} \textbf{\bibinfo{volume}{98}},
  \bibinfo{pages}{195103} (\bibinfo{year}{2018}).

\bibitem[{\citenamefont{Pogrebinskii}(1977)}]{pogrebinskii}
\bibinfo{author}{\bibfnamefont{M.}~\bibnamefont{Pogrebinskii}},
  \bibinfo{journal}{Fizika i Tekhnika Poluprovodnikov}
  \textbf{\bibinfo{volume}{11}}, \bibinfo{pages}{637} (\bibinfo{year}{1977}).

\bibitem[{\citenamefont{Price}(1983)}]{Price1983}
\bibinfo{author}{\bibfnamefont{P.~J.} \bibnamefont{Price}},
  \bibinfo{journal}{Physica B+C} \textbf{\bibinfo{volume}{117-118}},
  \bibinfo{pages}{750 } (\bibinfo{year}{1983}), ISSN \bibinfo{issn}{0378-4363}.

\bibitem[{\citenamefont{Price}()}]{Price1988}
\bibinfo{author}{\bibfnamefont{P.~J.} \bibnamefont{Price}},
  \bibinfo{howpublished}{in The Physics of Submicron Semiconductor Devices,
  edited by H. Grubin, D. K. Ferry, and C. Jacobon(1988)}.

\bibitem[{\citenamefont{Gramila et~al.}(1991)\citenamefont{Gramila, Eisenstein,
  MacDonald, Pfeiffer, and West}}]{Gramila}
\bibinfo{author}{\bibfnamefont{T.~J.} \bibnamefont{Gramila}},
  \bibinfo{author}{\bibfnamefont{J.~P.} \bibnamefont{Eisenstein}},
  \bibinfo{author}{\bibfnamefont{A.~H.} \bibnamefont{MacDonald}},
  \bibinfo{author}{\bibfnamefont{L.~N.} \bibnamefont{Pfeiffer}},
  \bibnamefont{and} \bibinfo{author}{\bibfnamefont{K.~W.} \bibnamefont{West}},
  \bibinfo{journal}{Phys. Rev. Lett.} \textbf{\bibinfo{volume}{66}},
  \bibinfo{pages}{1216} (\bibinfo{year}{1991}).

\bibitem[{\citenamefont{Zheng and MacDonald}(1993)}]{MacDonald}
\bibinfo{author}{\bibfnamefont{L.}~\bibnamefont{Zheng}} \bibnamefont{and}
  \bibinfo{author}{\bibfnamefont{A.~H.} \bibnamefont{MacDonald}},
  \bibinfo{journal}{Phys. Rev. B} \textbf{\bibinfo{volume}{48}},
  \bibinfo{pages}{8203} (\bibinfo{year}{1993}).

\bibitem[{\citenamefont{Flensberg and Hu}(1994)}]{Flensberg}
\bibinfo{author}{\bibfnamefont{K.}~\bibnamefont{Flensberg}} \bibnamefont{and}
  \bibinfo{author}{\bibfnamefont{B.~Y.-K.} \bibnamefont{Hu}},
  \bibinfo{journal}{Phys. Rev. Lett.} \textbf{\bibinfo{volume}{73}},
  \bibinfo{pages}{3572} (\bibinfo{year}{1994}).

\bibitem[{\citenamefont{Kamenev and Oreg}(1995)}]{Kamenev}
\bibinfo{author}{\bibfnamefont{A.}~\bibnamefont{Kamenev}} \bibnamefont{and}
  \bibinfo{author}{\bibfnamefont{Y.}~\bibnamefont{Oreg}},
  \bibinfo{journal}{Phys. Rev. B} \textbf{\bibinfo{volume}{52}},
  \bibinfo{pages}{7516} (\bibinfo{year}{1995}).

\bibitem[{\citenamefont{Solomon and Laikhtman}(1991)}]{Solomon}
\bibinfo{author}{\bibfnamefont{P.}~\bibnamefont{Solomon}} \bibnamefont{and}
  \bibinfo{author}{\bibfnamefont{B.}~\bibnamefont{Laikhtman}},
  \bibinfo{journal}{Superlattices and Microstructures}
  \textbf{\bibinfo{volume}{10}}, \bibinfo{pages}{89 } (\bibinfo{year}{1991}),
  ISSN \bibinfo{issn}{0749-6036}.

\bibitem[{\citenamefont{Jauho and Smith}(1993)}]{Jauho}
\bibinfo{author}{\bibfnamefont{A.-P.} \bibnamefont{Jauho}} \bibnamefont{and}
  \bibinfo{author}{\bibfnamefont{H.}~\bibnamefont{Smith}},
  \bibinfo{journal}{Phys. Rev. B} \textbf{\bibinfo{volume}{47}},
  \bibinfo{pages}{4420} (\bibinfo{year}{1993}).

\bibitem[{\citenamefont{Chen et~al.}(2015)\citenamefont{Chen, Andreev, and
  Levchenko}}]{Langevin}
\bibinfo{author}{\bibfnamefont{W.}~\bibnamefont{Chen}},
  \bibinfo{author}{\bibfnamefont{A.~V.} \bibnamefont{Andreev}},
  \bibnamefont{and}
  \bibinfo{author}{\bibfnamefont{A.}~\bibnamefont{Levchenko}},
  \bibinfo{journal}{Phys. Rev. B} \textbf{\bibinfo{volume}{91}},
  \bibinfo{pages}{245405} (\bibinfo{year}{2015}).

\bibitem[{\citenamefont{Flensberg and Hu}(1995)}]{Flensberg2}
\bibinfo{author}{\bibfnamefont{K.}~\bibnamefont{Flensberg}} \bibnamefont{and}
  \bibinfo{author}{\bibfnamefont{B.~Y.-K.} \bibnamefont{Hu}},
  \bibinfo{journal}{Phys. Rev. B} \textbf{\bibinfo{volume}{52}},
  \bibinfo{pages}{14796} (\bibinfo{year}{1995}).

\bibitem[{\citenamefont{Hill et~al.}(1997)\citenamefont{Hill, Nicholls,
  Linfield, Pepper, Ritchie, Jones, Hu, and Flensberg}}]{plasmon}
\bibinfo{author}{\bibfnamefont{N.~P.~R.} \bibnamefont{Hill}},
  \bibinfo{author}{\bibfnamefont{J.~T.} \bibnamefont{Nicholls}},
  \bibinfo{author}{\bibfnamefont{E.~H.} \bibnamefont{Linfield}},
  \bibinfo{author}{\bibfnamefont{M.}~\bibnamefont{Pepper}},
  \bibinfo{author}{\bibfnamefont{D.~A.} \bibnamefont{Ritchie}},
  \bibinfo{author}{\bibfnamefont{G.~A.~C.} \bibnamefont{Jones}},
  \bibinfo{author}{\bibfnamefont{B.~Y.-K.} \bibnamefont{Hu}}, \bibnamefont{and}
  \bibinfo{author}{\bibfnamefont{K.}~\bibnamefont{Flensberg}},
  \bibinfo{journal}{Physical review letters} \textbf{\bibinfo{volume}{78}},
  \bibinfo{pages}{2204} (\bibinfo{year}{1997}).

\bibitem[{\citenamefont{Badalyan et~al.}(2008)\citenamefont{Badalyan, Kim, and
  Vignale}}]{Width}
\bibinfo{author}{\bibfnamefont{S.~M.} \bibnamefont{Badalyan}},
  \bibinfo{author}{\bibfnamefont{C.~S.} \bibnamefont{Kim}}, \bibnamefont{and}
  \bibinfo{author}{\bibfnamefont{G.}~\bibnamefont{Vignale}},
  \bibinfo{journal}{Phys. Rev. Lett.} \textbf{\bibinfo{volume}{100}},
  \bibinfo{pages}{016603} (\bibinfo{year}{2008}).

\bibitem[{\citenamefont{Badalyan et~al.}(2007)\citenamefont{Badalyan, Kim,
  Vignale, and Senatore}}]{XC}
\bibinfo{author}{\bibfnamefont{S.~M.} \bibnamefont{Badalyan}},
  \bibinfo{author}{\bibfnamefont{C.~S.} \bibnamefont{Kim}},
  \bibinfo{author}{\bibfnamefont{G.}~\bibnamefont{Vignale}}, \bibnamefont{and}
  \bibinfo{author}{\bibfnamefont{G.}~\bibnamefont{Senatore}},
  \bibinfo{journal}{Phys. Rev. B} \textbf{\bibinfo{volume}{75}},
  \bibinfo{pages}{125321} (\bibinfo{year}{2007}).

\bibitem[{\citenamefont{Narozhny and Levchenko}(2016)}]{DragRev}
\bibinfo{author}{\bibfnamefont{B.~N.} \bibnamefont{Narozhny}} \bibnamefont{and}
  \bibinfo{author}{\bibfnamefont{A.}~\bibnamefont{Levchenko}},
  \bibinfo{journal}{Reviews of Modern Physics} \textbf{\bibinfo{volume}{88}},
  \bibinfo{pages}{025003} (\bibinfo{year}{2016}).

\bibitem[{\citenamefont{Rojo}(1999)}]{Rojo}
\bibinfo{author}{\bibfnamefont{A.~G.} \bibnamefont{Rojo}},
  \bibinfo{journal}{Journal of Physics: Condensed Matter}
  \textbf{\bibinfo{volume}{11}}, \bibinfo{pages}{R31} (\bibinfo{year}{1999}).

\bibitem[{\citenamefont{Stern}(1967)}]{stern}
\bibinfo{author}{\bibfnamefont{F.}~\bibnamefont{Stern}},
  \bibinfo{journal}{Physical Review Letters} \textbf{\bibinfo{volume}{18}},
  \bibinfo{pages}{546} (\bibinfo{year}{1967}).

\bibitem[{\citenamefont{Lindhard}(1954)}]{lindhard}
\bibinfo{author}{\bibfnamefont{J.}~\bibnamefont{Lindhard}},
  \bibinfo{journal}{Dan. Vid. Selsk Mat.-Fys. Medd.}
  \textbf{\bibinfo{volume}{28}}, \bibinfo{pages}{8} (\bibinfo{year}{1954}).

\bibitem[{\citenamefont{Kadanoff and Martin}(1963)}]{Kadanoff}
\bibinfo{author}{\bibfnamefont{L.~P.} \bibnamefont{Kadanoff}} \bibnamefont{and}
  \bibinfo{author}{\bibfnamefont{P.~C.} \bibnamefont{Martin}},
  \bibinfo{journal}{Annals of Physics} \textbf{\bibinfo{volume}{24}},
  \bibinfo{pages}{419} (\bibinfo{year}{1963}).

\bibitem[{\citenamefont{Bradlyn et~al.}(2012)\citenamefont{Bradlyn, Goldstein,
  and Read}}]{Read}
\bibinfo{author}{\bibfnamefont{B.}~\bibnamefont{Bradlyn}},
  \bibinfo{author}{\bibfnamefont{M.}~\bibnamefont{Goldstein}},
  \bibnamefont{and} \bibinfo{author}{\bibfnamefont{N.}~\bibnamefont{Read}},
  \bibinfo{journal}{Physical Review B} \textbf{\bibinfo{volume}{86}},
  \bibinfo{pages}{245309} (\bibinfo{year}{2012}).

\bibitem[{\citenamefont{Forster}(2018)}]{forster}
\bibinfo{author}{\bibfnamefont{D.}~\bibnamefont{Forster}},
  \emph{\bibinfo{title}{Hydrodynamic fluctuations, broken symmetry, and
  correlation functions}} (\bibinfo{publisher}{CRC Press},
  \bibinfo{year}{2018}).

\bibitem[{\citenamefont{Zubarev and Shepherd}(1974)}]{zubarev}
\bibinfo{author}{\bibfnamefont{D.~N.} \bibnamefont{Zubarev}} \bibnamefont{and}
  \bibinfo{author}{\bibfnamefont{P.}~\bibnamefont{Shepherd}},
  \emph{\bibinfo{title}{Nonequilibrium statistical thermodynamics}}
  (\bibinfo{publisher}{Consultants Bureau New York}, \bibinfo{year}{1974}).

\bibitem[{\citenamefont{Martin and Schwinger}(1959)}]{Schwinger}
\bibinfo{author}{\bibfnamefont{P.~C.} \bibnamefont{Martin}} \bibnamefont{and}
  \bibinfo{author}{\bibfnamefont{J.}~\bibnamefont{Schwinger}},
  \bibinfo{journal}{Physical Review} \textbf{\bibinfo{volume}{115}},
  \bibinfo{pages}{1342} (\bibinfo{year}{1959}).

\bibitem[{\citenamefont{Tao et~al.}(2017)\citenamefont{Tao, Vignale, and
  Zhu}}]{geoT}
\bibinfo{author}{\bibfnamefont{J.}~\bibnamefont{Tao}},
  \bibinfo{author}{\bibfnamefont{G.}~\bibnamefont{Vignale}}, \bibnamefont{and}
  \bibinfo{author}{\bibfnamefont{J.-X.} \bibnamefont{Zhu}},
  \bibinfo{journal}{Computation} \textbf{\bibinfo{volume}{5}},
  \bibinfo{pages}{28} (\bibinfo{year}{2017}).

\bibitem[{\citenamefont{Zyuzin et~al.}(2018)\citenamefont{Zyuzin, Sharma, and
  Maslov}}]{Maslov}
\bibinfo{author}{\bibfnamefont{V.~A.} \bibnamefont{Zyuzin}},
  \bibinfo{author}{\bibfnamefont{P.}~\bibnamefont{Sharma}}, \bibnamefont{and}
  \bibinfo{author}{\bibfnamefont{D.~L.} \bibnamefont{Maslov}},
  \bibinfo{journal}{Phys. Rev. B} \textbf{\bibinfo{volume}{98}},
  \bibinfo{pages}{115139} (\bibinfo{year}{2018}).

\bibitem[{\citenamefont{Conti and Vignale}(1999)}]{Conti}
\bibinfo{author}{\bibfnamefont{S.}~\bibnamefont{Conti}} \bibnamefont{and}
  \bibinfo{author}{\bibfnamefont{G.}~\bibnamefont{Vignale}},
  \bibinfo{journal}{Phys. Rev. B} \textbf{\bibinfo{volume}{60}},
  \bibinfo{pages}{7966} (\bibinfo{year}{1999}).

\bibitem[{\citenamefont{Tse et~al.}(2007)\citenamefont{Tse, Hu, and {Das
  Sarma}}}]{GDrag-DasSarma}
\bibinfo{author}{\bibfnamefont{W.-K.} \bibnamefont{Tse}},
  \bibinfo{author}{\bibfnamefont{B.~Y.-K.} \bibnamefont{Hu}}, \bibnamefont{and}
  \bibinfo{author}{\bibfnamefont{S.}~\bibnamefont{{Das Sarma}}},
  \bibinfo{journal}{Phys. Rev. B} \textbf{\bibinfo{volume}{76}},
  \bibinfo{pages}{081401(R)} (\bibinfo{year}{2007}).

\bibitem[{\citenamefont{Narozhny}(2007)}]{GDrag-Narozhny}
\bibinfo{author}{\bibfnamefont{B.~N.} \bibnamefont{Narozhny}},
  \bibinfo{journal}{Phys. Rev. B} \textbf{\bibinfo{volume}{76}},
  \bibinfo{pages}{153409} (\bibinfo{year}{2007}).

\bibitem[{\citenamefont{Song and Levitov}(2013)}]{Mdrag-Levitov}
\bibinfo{author}{\bibfnamefont{J.~C.~W.} \bibnamefont{Song}} \bibnamefont{and}
  \bibinfo{author}{\bibfnamefont{L.~S.} \bibnamefont{Levitov}},
  \bibinfo{journal}{Phys. Rev. Lett.} \textbf{\bibinfo{volume}{111}},
  \bibinfo{pages}{126601} (\bibinfo{year}{2013}).

\bibitem[{\citenamefont{Titov et~al.}(2013)\citenamefont{Titov, Gorbachev,
  Narozhny, Tudorovskiy, Sch\"utt, Ostrovsky, Gornyi, Mirlin, Katsnelson,
  Novoselov et~al.}}]{Mdrag-Titov}
\bibinfo{author}{\bibfnamefont{M.}~\bibnamefont{Titov}},
  \bibinfo{author}{\bibfnamefont{R.~V.} \bibnamefont{Gorbachev}},
  \bibinfo{author}{\bibfnamefont{B.~N.} \bibnamefont{Narozhny}},
  \bibinfo{author}{\bibfnamefont{T.}~\bibnamefont{Tudorovskiy}},
  \bibinfo{author}{\bibfnamefont{M.}~\bibnamefont{Sch\"utt}},
  \bibinfo{author}{\bibfnamefont{P.~M.} \bibnamefont{Ostrovsky}},
  \bibinfo{author}{\bibfnamefont{I.~V.} \bibnamefont{Gornyi}},
  \bibinfo{author}{\bibfnamefont{A.~D.} \bibnamefont{Mirlin}},
  \bibinfo{author}{\bibfnamefont{M.~I.} \bibnamefont{Katsnelson}},
  \bibinfo{author}{\bibfnamefont{K.~S.} \bibnamefont{Novoselov}},
  \bibnamefont{et~al.}, \bibinfo{journal}{Phys. Rev. Lett.}
  \textbf{\bibinfo{volume}{111}}, \bibinfo{pages}{166601}
  (\bibinfo{year}{2013}).

\bibitem[{\citenamefont{Gorbachev et~al.}(2012)\citenamefont{Gorbachev, Geim,
  Katsnelson, Novoselov, Tudorovskiy, Grigorieva, MacDonald, Morozov, Watanabe,
  Taniguchi et~al.}}]{Mdrag-exp}
\bibinfo{author}{\bibfnamefont{R.}~\bibnamefont{Gorbachev}},
  \bibinfo{author}{\bibfnamefont{A.}~\bibnamefont{Geim}},
  \bibinfo{author}{\bibfnamefont{M.}~\bibnamefont{Katsnelson}},
  \bibinfo{author}{\bibfnamefont{K.}~\bibnamefont{Novoselov}},
  \bibinfo{author}{\bibfnamefont{T.}~\bibnamefont{Tudorovskiy}},
  \bibinfo{author}{\bibfnamefont{I.}~\bibnamefont{Grigorieva}},
  \bibinfo{author}{\bibfnamefont{A.~H.} \bibnamefont{MacDonald}},
  \bibinfo{author}{\bibfnamefont{S.}~\bibnamefont{Morozov}},
  \bibinfo{author}{\bibfnamefont{K.}~\bibnamefont{Watanabe}},
  \bibinfo{author}{\bibfnamefont{T.}~\bibnamefont{Taniguchi}},
  \bibnamefont{et~al.}, \bibinfo{journal}{Nature Physics}
  \textbf{\bibinfo{volume}{8}}, \bibinfo{pages}{896} (\bibinfo{year}{2012}).

\bibitem[{\citenamefont{Eberlein et~al.}(2017)\citenamefont{Eberlein, Patel,
  and Sachdev}}]{QCP}
\bibinfo{author}{\bibfnamefont{A.}~\bibnamefont{Eberlein}},
  \bibinfo{author}{\bibfnamefont{A.~A.} \bibnamefont{Patel}}, \bibnamefont{and}
  \bibinfo{author}{\bibfnamefont{S.}~\bibnamefont{Sachdev}},
  \bibinfo{journal}{Phys. Rev. B} \textbf{\bibinfo{volume}{95}},
  \bibinfo{pages}{075127} (\bibinfo{year}{2017}).

\bibitem[{\citenamefont{Varma et~al.}(1989)\citenamefont{Varma, Littlewood,
  Schmitt-Rink, Abrahams, and Ruckenstein}}]{cuprate}
\bibinfo{author}{\bibfnamefont{C.~M.} \bibnamefont{Varma}},
  \bibinfo{author}{\bibfnamefont{P.~B.} \bibnamefont{Littlewood}},
  \bibinfo{author}{\bibfnamefont{S.}~\bibnamefont{Schmitt-Rink}},
  \bibinfo{author}{\bibfnamefont{E.}~\bibnamefont{Abrahams}}, \bibnamefont{and}
  \bibinfo{author}{\bibfnamefont{A.~E.} \bibnamefont{Ruckenstein}},
  \bibinfo{journal}{Phys. Rev. Lett.} \textbf{\bibinfo{volume}{63}},
  \bibinfo{pages}{1996} (\bibinfo{year}{1989}).

\bibitem[{\citenamefont{Davison et~al.}(2014)\citenamefont{Davison, Schalm, and
  Zaanen}}]{Davison}
\bibinfo{author}{\bibfnamefont{R.~A.} \bibnamefont{Davison}},
  \bibinfo{author}{\bibfnamefont{K.}~\bibnamefont{Schalm}}, \bibnamefont{and}
  \bibinfo{author}{\bibfnamefont{J.}~\bibnamefont{Zaanen}},
  \bibinfo{journal}{Phys. Rev. B} \textbf{\bibinfo{volume}{89}},
  \bibinfo{pages}{245116} (\bibinfo{year}{2014}).

\bibitem[{\citenamefont{Spivak and Kivelson}(2006)}]{Spivak}
\bibinfo{author}{\bibfnamefont{B.}~\bibnamefont{Spivak}} \bibnamefont{and}
  \bibinfo{author}{\bibfnamefont{S.~A.} \bibnamefont{Kivelson}},
  \bibinfo{journal}{Annals of Physics} \textbf{\bibinfo{volume}{321}},
  \bibinfo{pages}{2071 } (\bibinfo{year}{2006}), ISSN
  \bibinfo{issn}{0003-4916}.

\bibitem[{\citenamefont{Loram et~al.}(1993)\citenamefont{Loram, Mirza, Cooper,
  and Liang}}]{EntropyCuprate}
\bibinfo{author}{\bibfnamefont{J.~W.} \bibnamefont{Loram}},
  \bibinfo{author}{\bibfnamefont{K.~A.} \bibnamefont{Mirza}},
  \bibinfo{author}{\bibfnamefont{J.~R.} \bibnamefont{Cooper}},
  \bibnamefont{and} \bibinfo{author}{\bibfnamefont{W.~Y.} \bibnamefont{Liang}},
  \bibinfo{journal}{Phys. Rev. Lett.} \textbf{\bibinfo{volume}{71}},
  \bibinfo{pages}{1740} (\bibinfo{year}{1993}).

\end{thebibliography}

\end{document}